 \documentclass[12pt,preprint]{aastex}

 \newcommand{\kms}{~km s$^{-1}$~}
\newcommand{\myemail}{zhekovs@colorado.edu}

\shorttitle{LETG Observations of SNR~1987A} 
\shortauthors{Zhekov et al.}

\begin{document}


\title{Chandra LETG Observations of Supernova Remnant 1987A}


\author{Svetozar A. Zhekov\altaffilmark{1,4},
Richard McCray\altaffilmark{1},
Kazimierz J. Borkowski\altaffilmark{2},
David N. Burrows \altaffilmark{3}, and Sangwook Park\altaffilmark{3}}

%

\altaffiltext{1}{JILA, University of Colorado, Boulder, CO 80309-0440;
\myemail, dick@jila.colorado.edu}
\altaffiltext{2}{Department of Physics, NCSU, Raleigh, NC 27695-8202;
kborkow@unity.ncsu.edu}
\altaffiltext{3}{Department of Astronomy and Astrophysics,
Pennsylvania
State University, 525 Davey Laboratory, University Park, PA 16802;
burrows@astro.psu.edu; park@astro.psu.edu}
\altaffiltext{4}{Current address: Space Research Institute, 
Moskovska str. 6, Sofia-1000, Bulgaria}


\begin{abstract}   
We discuss the results from deep {\it Chandra} LETG
observations of the supernova remnant 1987A (SNR~1987A). We find that
a distribution of shocks, spanning the same range of velocities 
(from $\sim 300 \mbox{ to } 1700$\kms) as
deduced in the first part of our analysis (Zhekov et al. 2005, \apjl,
628, L127), can account for the entire X-ray spectrum of this object.
The post-shock temperature distribution is {\it bimodal}, peaking at
kT$\sim 0.5$~and $\sim 3$~keV.  Abundances inferred from the X-ray
spectrum have values similar to those for the inner
circumstellar ring, except that the abundances of nitrogen and oxygen
are approximately a factor of two lower than those inferred from the
optical/UV spectrum.  The velocity of the X-ray emitting plasma has
decreased since 1999, apparently because the blast wave has entered
the main body of the inner circumstellar ring.  
\end{abstract}





\section{Introduction} 
\label{sec:intro} 
With the rapidly developing
impact of the debris of Supernova 1987A with its inner circumstellar
ring, we have an unprecedented opportunity to witness the birth of a
supernova remnant, SNR~1987A (McCray 2005).  The reappearance of
SNR~1987A in radio (Staveley-Smith et al. 1992, 1993) and in X-rays
(Beuermann, Brandt \& Pietsch 1994; Gorenstein, Huges \& Tucker 1994;
Hasinger, Aschenbach \& Tr\"umper 1996) at $\sim 1200$~days after the
explosion was the first clear sign of this phenomenon. The presence of
a triple ring system of relatively dense circumstellar matter (CSM)
centered on the supernova (Burrows et al. 1995) ensures that
astronomers will be able to follow this exciting event for a long
time.  Indeed, we see the continuous brightening in radio (Gaensler et
al. 1997; Manchester et al. 2003) and X-rays (Park et al. 2004; 2005)
as well as the appearance in the optical and ultraviolet (UV) of
several rapidly brightening spatially unresolved hot spots (Lawrence
et al. 2000; Michael et al. 2000), which by now encircle the entire
inner ring (Sugerman et al. 2002).

The hydrodynamics of the interaction of the supernova debris with its
circumstellar matter is quite complex (Chevalier, Blondin \& Emmering
1992; Borkowski, Blondin \& McCray 1997a,b) and it depends on the
details of the density distribution of the CSM as well as the density
and velocity distributions in the supernova debris.  In general, a
double shock structure forms. The forward shock (blast wave)
propagates into the CSM, while a reverse shock propagates backwards
into the supernova debris.  Typically, the separation between the
blast wave and the reverse shock is $\sim 10 \%$ of the radius of the
blast wave (Chevalier 1982).  Between these two shocks are layers of
shocked CSM and shocked supernova debris, separated by a (usually
unstable) contact discontinuity.  These layers are sources of X-ray
emission. 

Hydrodynamic simulations show that, during the first decades of the
supernova remnant evolution, the total soft X-ray emission will be
dominated by the shocked CSM behind the blast wave (Borkowski et al.
1997a).  Then, when the blast wave strikes regions of dense gas in the
inner circumstellar ring, slower shocks will be transmitted into the
ring, while reflected shocks will propagate backwards through the
shocked CSM and will eventually merge with the reverse shock.  This
interaction will further enhance the X-ray emission.  By analyzing the
evolution of the X-ray spectrum of SNR~1987A, we have a unique
opportunity to learn about the density distribution and elemental
abundances of the supernova CSM, as well as the physics of high
velocity shocks in rarefied gases. These studies may also provide
vital clues to origin of the inner ring. 

Continuous monitoring of {\it Chandra} images of SNR~1987A (e.g.
Burrows et al. 2000; Park et al. 2002, 2004, 2005; Park et al. 2006,
in preparation) have provided us
with much information about the evolution of the spatial and thermal
distribution of the shocked gas.  
However, the poor spectral resolution of the CCD data do not
allow us to study the kinematics of the X-ray emitting gas in detail.  
Such information can be obtained only by analyzing 
high-resolution dispersed X-ray spectra.  

The first dispersed X-ray spectrum of SNR~1987A, obtained in October
1999 with the {\it Chandra} HETG, was discussed by Michael et al.
(2002).  Unfortunately, the limited photon statistics only allowed
these authors to construct a composite line profile.  The $\sim
5000$\kms FWHM of this profile indicated a blast wave (forward shock)
velocity $\sim 3500$\kms. When we obtained new {\it Chandra} (LETG)
observations in September 2004, we expected to see comparable
kinematic velocities; but we were surprised to find that the X-ray
emission lines had considerably smaller intrinsic widths ($\sim 500 -
1000$\kms).  This observation showed clearly that the X-ray emission
is now dominated by gas behind shocks of much lower velocities than
before (Zhekov et al. 2005; hereafter Paper I).

In Paper I, we measured the widths, line shifts, and fluxes of the
strong emission lines from hydrogen- and helium-like ions of abundant
chemical elements in the spectral energy range from 0.4 to 3~keV. (At
higher energies the spectrum is dominated by continuum emission.) We
inferred that the X-ray emission from this object is likely to
originate from a distribution of shocks with velocities in the range
$300-1700$\kms. We proposed that this ensemble of shocks is a
consequence of the blast wave interaction with the complex density
distribution of the inner circumstellar ring, where transmitted as
well as reflected shocks play a major role.

In this, the second part of our analysis of the {\it Chandra} LETG
observations, we test these conclusions further by carrying out a
`global' analysis.  In \S \ref{sec:obs} we review briefly our
observations and data reduction procedures.  In \S \ref{sec:lines} we
summarize the properties of strong X-ray emission lines as described
in Paper I. Then, in \S \ref{sec:global} and \S \ref{sec:marx} we
analyze the entire LETG spectrum of SNR~1987A and construct a model to
fit it.  Finally, in \S \ref{sec:disc} we discuss what our
observations imply about the overall picture of the newborn supernova
remnant SNR~1987A and its future evolution.  

\section{Observations and Data Reduction } 
\label{sec:obs} 
SNR~1987A
was observed with {\it Chandra} in the configuration LETG-ACIS-S in
five consecutive runs (\dataset[ADS/Sa.CXO\#obs/04640]{Chandra ObsId:
4640,} \dataset[ADS/Sa.CXO\#obs/04641]{ 4641,}
\dataset[ADS/Sa.CXO\#obs/05362]{ 5362,}
\dataset[ADS/Sa.CXO\#obs/05363]{ 5363 and}
\dataset[ADS/Sa.CXO\#obs/06099]{ 6099)} 
in the period Aug 26 -- Sep 5, 2004 ($\sim
6398$~days after the explosion), providing a total effective exposure
of 289 ksec. The roll angle was chosen so that the negative arm of the
dispersion axis was aligned approximately north, within $\sim
15^\circ$ from North to West, thus within $\sim 10^\circ$ of the minor
axis of the inner circumstellar ring (P.A. $\approx 354^\circ$
~Sugerman et al. 2002).

We extracted positive $(m = +1)$ and negative $(m = -1)$ first-order
LETG spectra for each of the five observations according to the
procedure described in the Science Threads for Grating Spectroscopy in
the CIAO 3.1 \,\footnote{Chandra Interactive Analysis of Observations
(CIAO), http://cxc.harvard.edu/ciao/} data analysis software.  We
merged the resultant spectra into one spectrum each for the positive
and negative LETG arms with respective total counts of 9,241 and 6,057
in the energy range 0.4 - 7 keV. The difference in photon statistics
is a result of the different sensitivities of the respective CCD
detectors.  For each of the five data sets we extracted a pulse-height
spectrum from the zeroth-order image (according to CIAO Science
Threads for Image Spectroscopy), contained in a circular region with a
radius of 10 pixels centered on the source position.  We measured the
corresponding background spectrum from an annulus with inner and outer
radii of 10 and 20 pixels, respectively.  After combining the five
pulse-height spectra into one, we measure a total number of 16,557
counts in the 0.4 - 7 keV range.  We generated the ancillary response
functions for all LETG spectra (1-st and 0-th order) using the Chandra
calibration database CALDB v3.00. 

\section{Spectral Lines} 
\label{sec:lines} 
\begin{figure} 
\begin{center} 
\includegraphics[width=3in, height=2.5in]{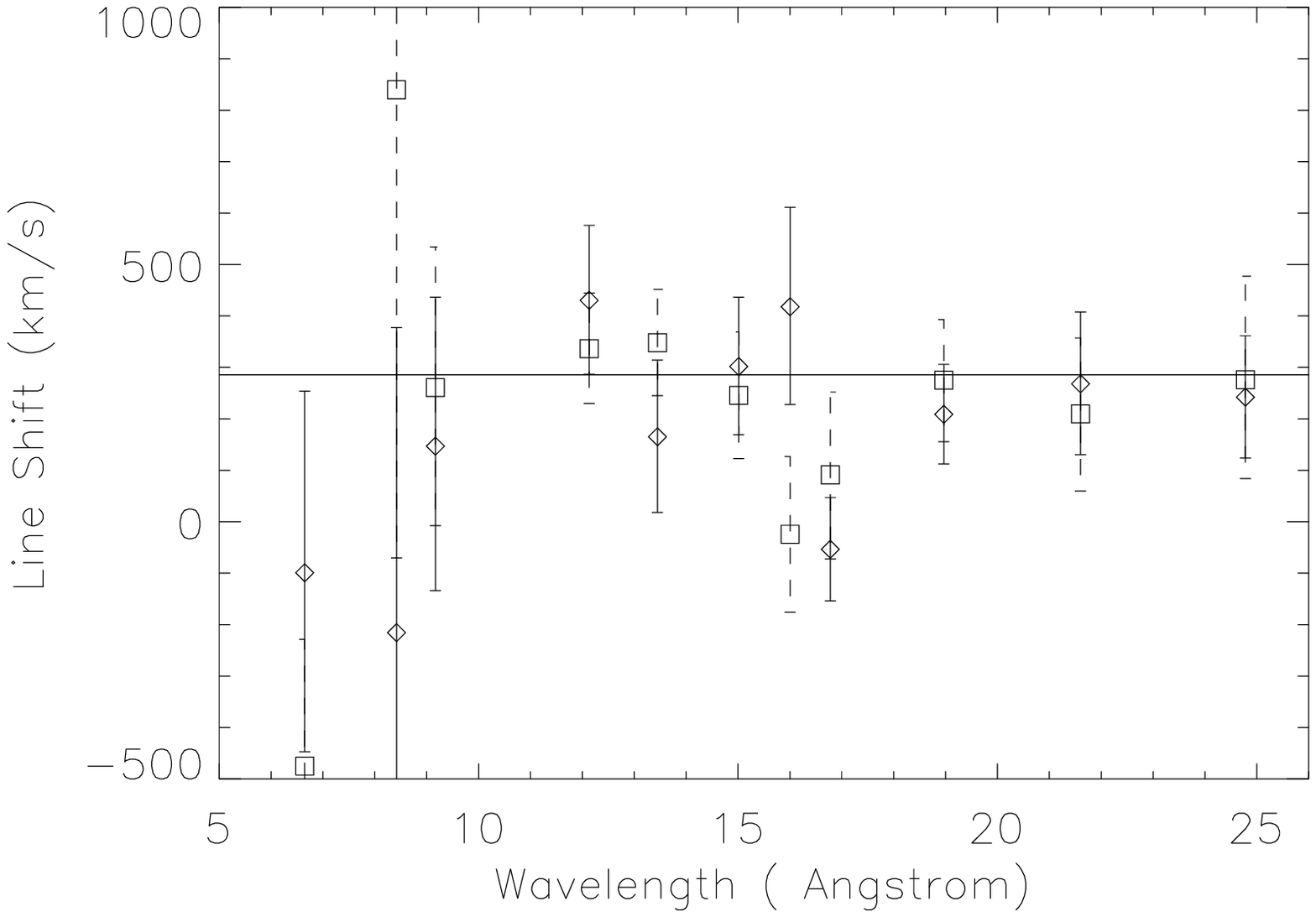} 
\includegraphics[width=2.5in, height=2.5in]{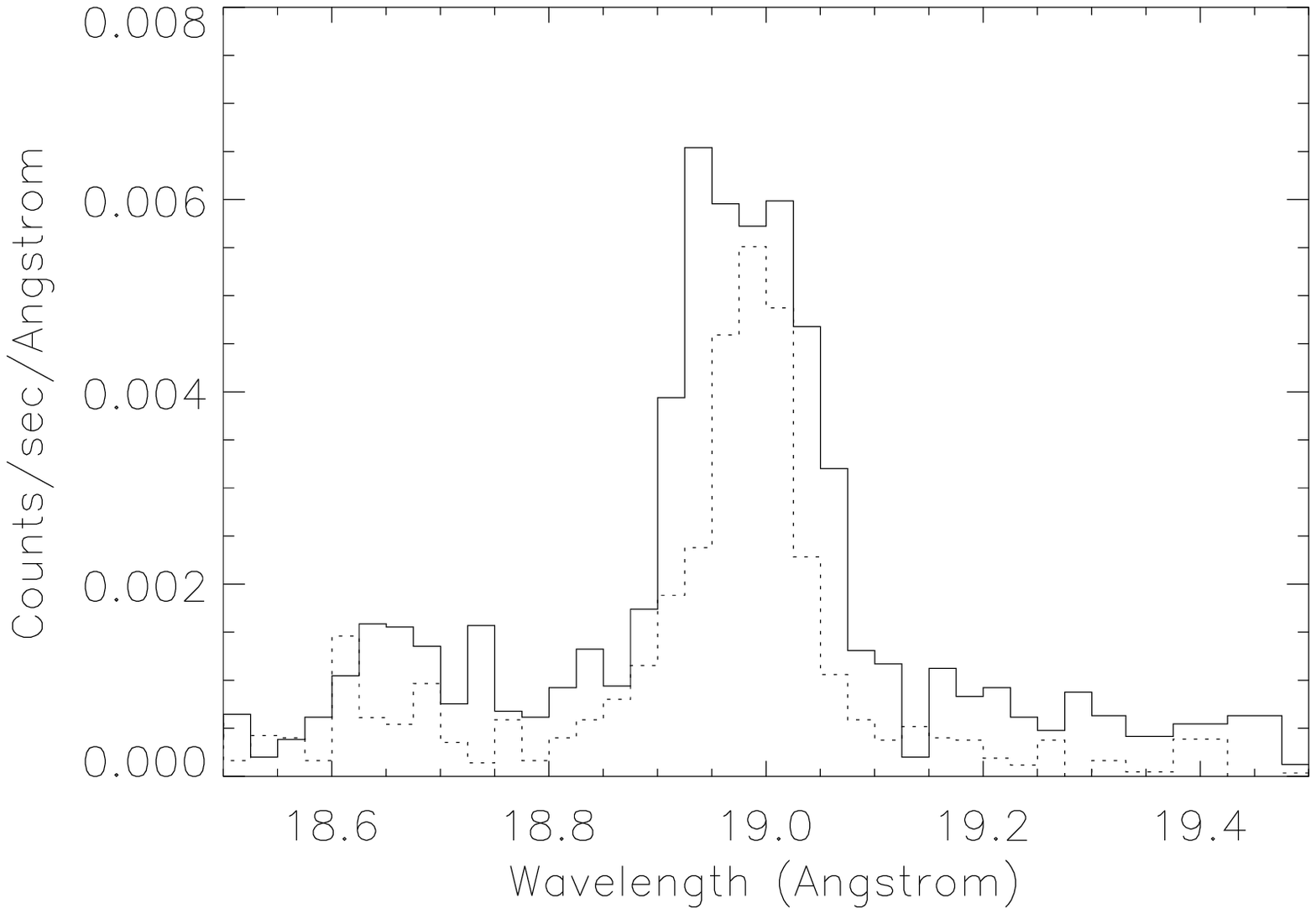} 
\end{center} 
\caption{ 
{\bf Left panel:} Line shifts for the strong emission lines in the positive
({\it diamonds}) and negative ({\it squares}) LETG first-order spectra
of SN 1987A. The solid line in the left panel represents the average
red-shift for the Large Magellanic Cloud (z $= 9.53\times10^{-4}$).
{\bf Right panel:} The OVIII L$_{\alpha}$~profile in the positive
(solid line) and negative (dashed line) LETG first-order spectra. The
count rate in the $m = -1$ spectrum has been multiplied by a factor
1.5 to correct for the lower sensitivity of the detector in that arm.
} 
\label{fig:shift} 
\end{figure}

Here we summarize the basic
results from our analysis in Paper I of the strong X-ray emission
lines in the spectrum.  We found that the line profiles had no
apparent asymmetries and that the centroids for the emission lines
were consistent with the redshift for the Large Magellanic Cloud (see
also Fig. \ref{fig:shift}).  Therefore, we could find satisfactory
Gaussian fits to the line profiles.  
All the lines seen in ($m = -1$) LETG
spectrum have smaller widths than their counterparts in the ($m = +1$)
spectrum.  As described in Paper I, this phenomenon results from the
convolution of the spatial structure and velocity gradients in the
X-ray source.\footnote{This effect is also discussed in The Chandra
Proposer's Observatory Guide, Version 7.0, \S 8.5.3, pp.187-189, \S
9.3.3, pp. 209-215.} It is a consequence of the facts that the X-ray
source is resolved and that the X-ray emitting plasma is confined in
the plane of an expanding ring.  
By having anticipated this physical picture and
to use this effect to maximum
advantage, we chose the roll angle for the LETG observations to
coincide nearly with the minor axis of the inner circumstellar ring.  

\section{Global Spectral Fits} 
\label{sec:global} 
A systematic
approach to modeling the X-ray spectrum of SNR~1987A would be to make
a global fit of the entire observed spectrum using a model consisting
of a distribution of shock velocities and ages.  Such a global fit has
significant advantages over the line ratio analysis presented in Paper
I: (1) the fit can automatically take into account the
quasi-continuum due to numerous weak lines; (2) by fitting the shape
of the underlying continuum (free-free, recombination, and two-photon
emission) the model places additional constraints on the plasma
temperature; (3) the model can constrain the column density of X-ray
absorbing gas; and (4) it can yield estimates of relative element
abundances.

It is prohibitive, however, to explore the entire range of models
defined by such a continuum in parameter space.  Instead, we begin by
exploring fits to the spectrum with a model consisting of one or two
plane-parallel shocks and then consider a more complex distribution of
shocks.

In adopting this approach, we have used the recent version (11.3.2) of
the XSPEC code for modeling X-ray emission spectra of shocks.
Specifically, we employ the {\it vpshock} model, which takes into
account the non-equilibrium ionization in hot plasmas 
from version 2.0 of NEI models in XSPEC, which are based on ATOMDB
(Smith et al. 2001). We augmented this atomic database by adding
inner-shell processes, which are missing in NEI v2.0.
In such shock models,
the X-ray emission properties are functions of the post-shock
temperature and the ionization age, $n_et$, defined as the product of
the postshock electron density and the time since the gas first
entered the shock.  Details of these models 
are found in Borkowski, Lyerly, \& Reynolds (2001).

To determine the elemental abundances in our fits, we adopted the same
procedure as in our previous spectral analyses (Michael et al. 2002;
Park et al. 2002, 2004). That is, we only varied the abundances of
elements having strong emission lines in the observed (0.5-4~keV)
energy range, namely N, O, Ne, Mg, Si, S and Fe.  We fixed the
abundances of He and C to the values determined by Lundqvist \&
Fransson (1996) from fits to the optical/UV spectrum of the inner ring
and the abundances of the remaining elements (Ar, Ca and Ni) 
to values representative
of the Large Magellanic Cloud (Russel \& Dopita 1992).

We fitted the positive ($m = + 1$) and negative ($m = - 1$) LETG 1-st
order spectra simultaneously with models having identical plasma
characteristics but different line broadening parameters defined as in
Paper I. Namely, we assumed that the line profiles are Gaussian with
three sources of line broadening: (1) the spatial extent of the image
itself, which can be expressed as an equivalent line broadening,
independent of wavelength; (2) the thermal broadening of the shocked
gas; and (3) the broadening due to the bulk motion of the shocked gas.
For plasma temperatures of interest here (kT $\approx 0.1 - 4$~keV)
the thermal line broadening for heavy elements (N through Fe) is
negligible compared to the other two sources.  
It is so since the standard relations between the shock parameters 
yield a ratio of the ion thermal velocity to the postshock bulk
gas velocity: 
$V_{th}/V_{bulk} = \sqrt{\frac{2}{3} \frac{\mu}{A}} = 0.7/\sqrt{A}$
~ where
the mean particle weight $\mu = 0.72$~for SNR 1987A and $A$ is the
mass of a given ion in units of the proton mass. Therefore,
the thermal velocity of various ionic species is only between 10\% and 20\%
of the bulk gas velocity ($A = 14 \mbox{ [N] through } 56 \mbox{ [Fe]}$).
On the other hand, the spatial size of SNR~1987A in wavelength units 
($\Delta \lambda_0 = 0.047$~\AA, Paper I)
results in a line broadening larger than that from the
bulk gas velocity of the shocked gas.
Therefore, we express
the net line width (FWHM) as: 
\begin{equation} 
\Delta \lambda_{tot} =
2 \Delta \lambda_0 \pm 2 z_0 (\lambda/\lambda_0)^\alpha \lambda\,,~
\label{eqn:stratified} 
\end{equation} 
where the plus (minus) sign
refers to the $m = + 1$ ($m = - 1$) spectrum, respectively. The first
term on the right hand side of equation~(\ref{eqn:stratified})
represents the broadening due to the spatial extent of the source. The
second term represents the broadening due to the bulk motion.  The
power-law function of wavelength in the second term allows for the
possibility that the mean bulk velocity of shocked gas emitting a
given line may depend on the excitation or ionization stage of the
emitting ion.  The parameter $z_0$ determines the line broadening at
some fiducial wavelength, $\lambda_0$, and the power-law index
$\alpha$ is to be determined.

\subsection{Discrete Shock Models}

\begin{figure}
\begin{center}
\includegraphics[width=3in, height=1.85in]{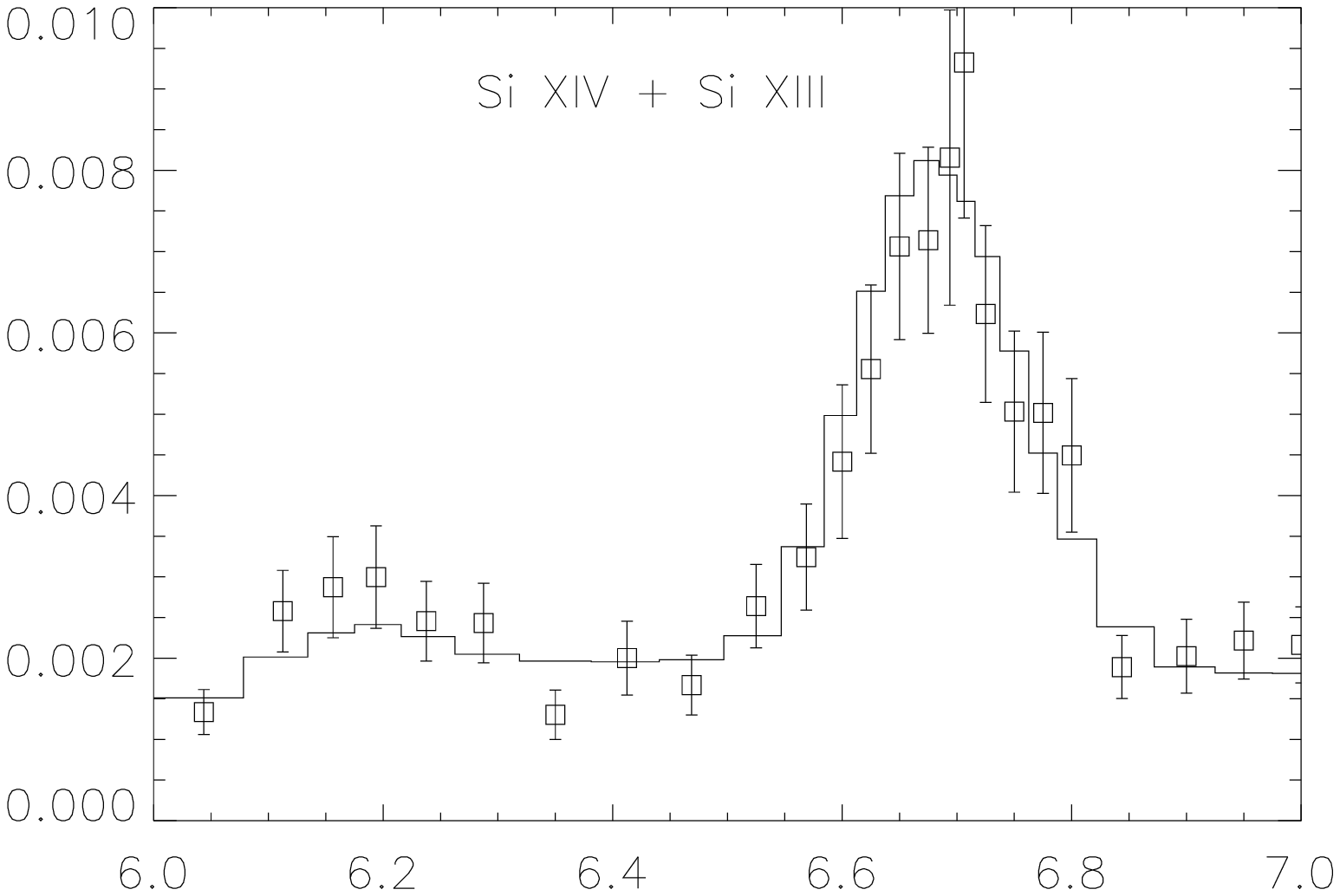}
\includegraphics[width=3in, height=1.85in]{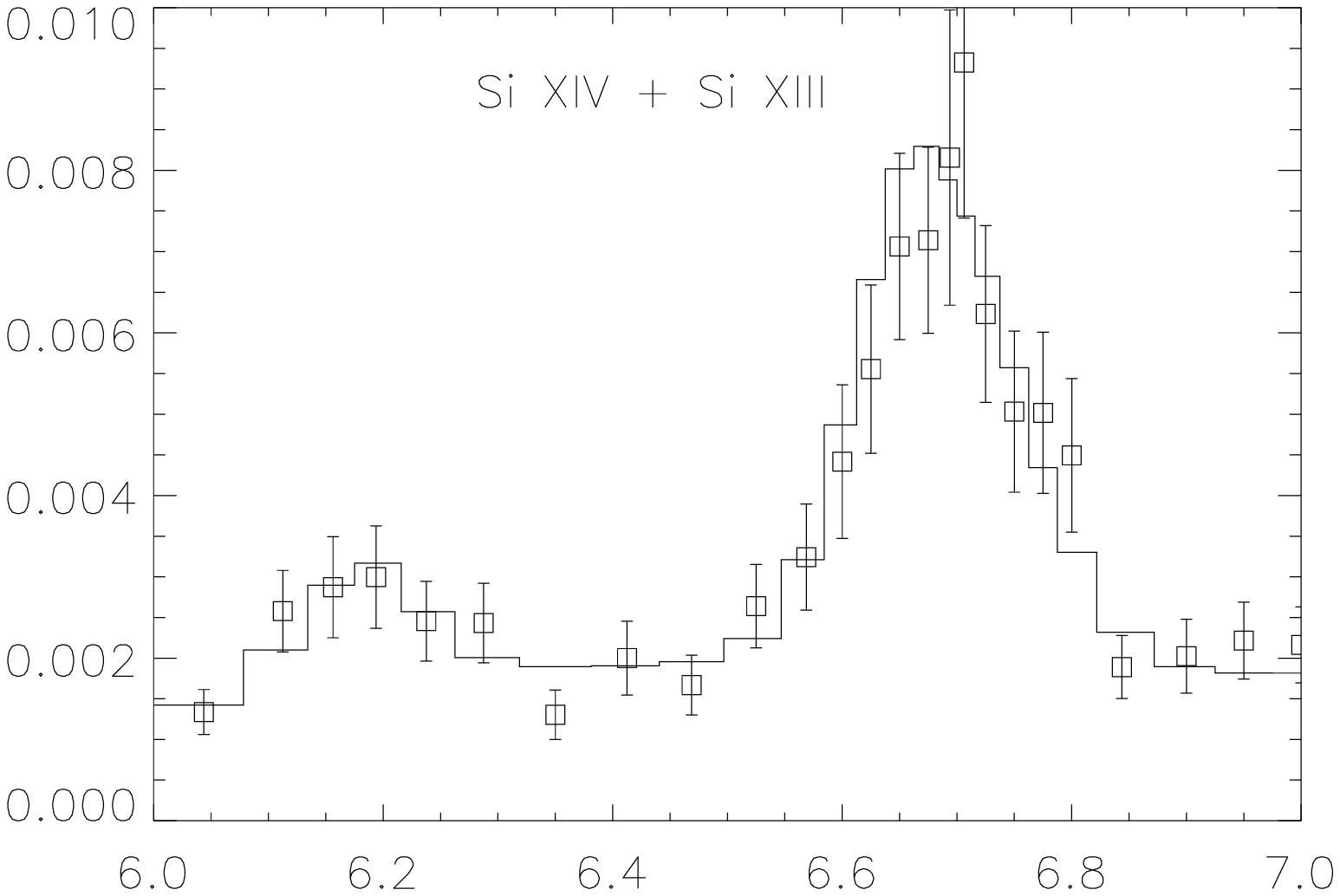}
\includegraphics[width=3in, height=1.85in]{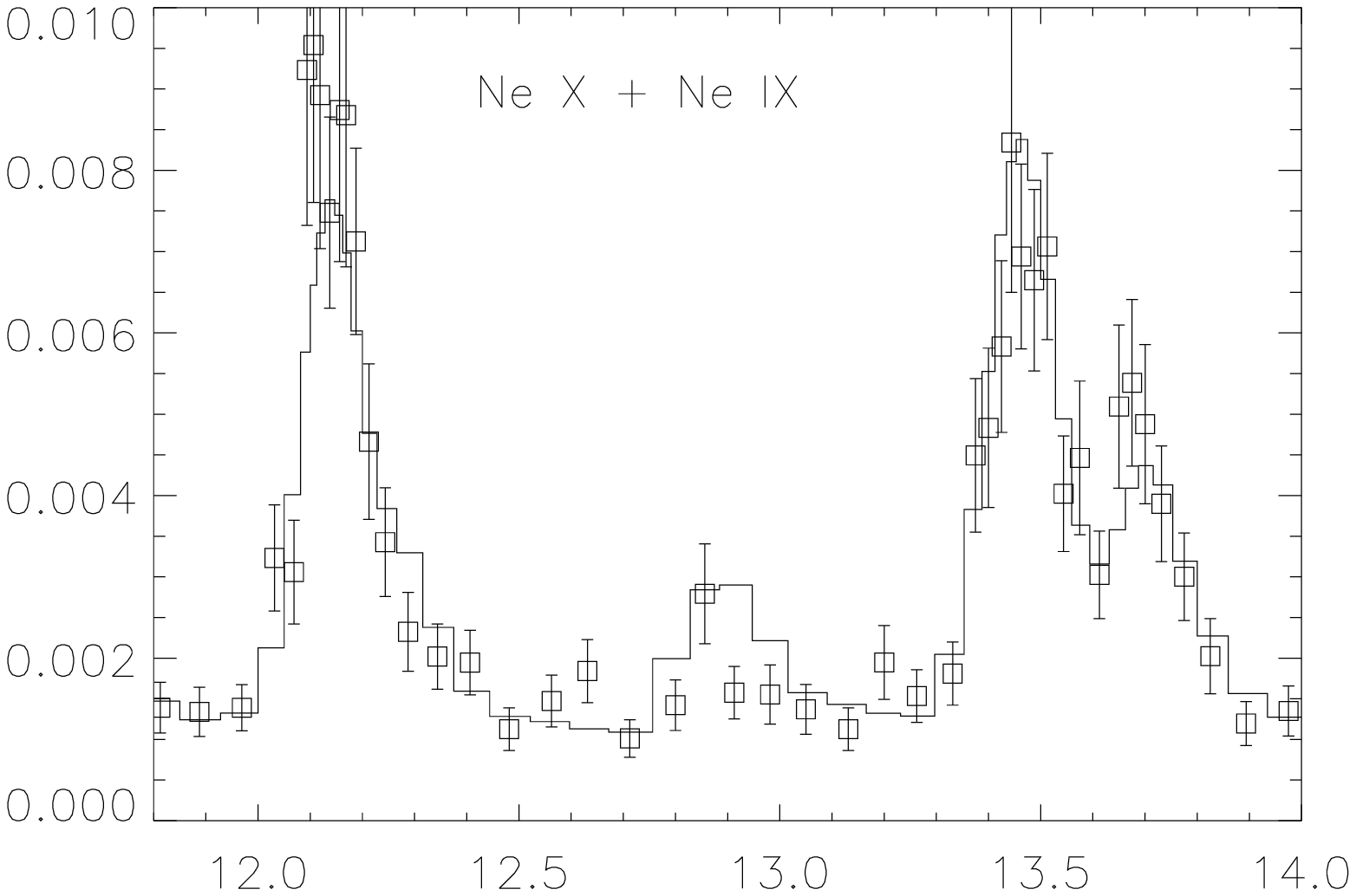}
\includegraphics[width=3in, height=1.85in]{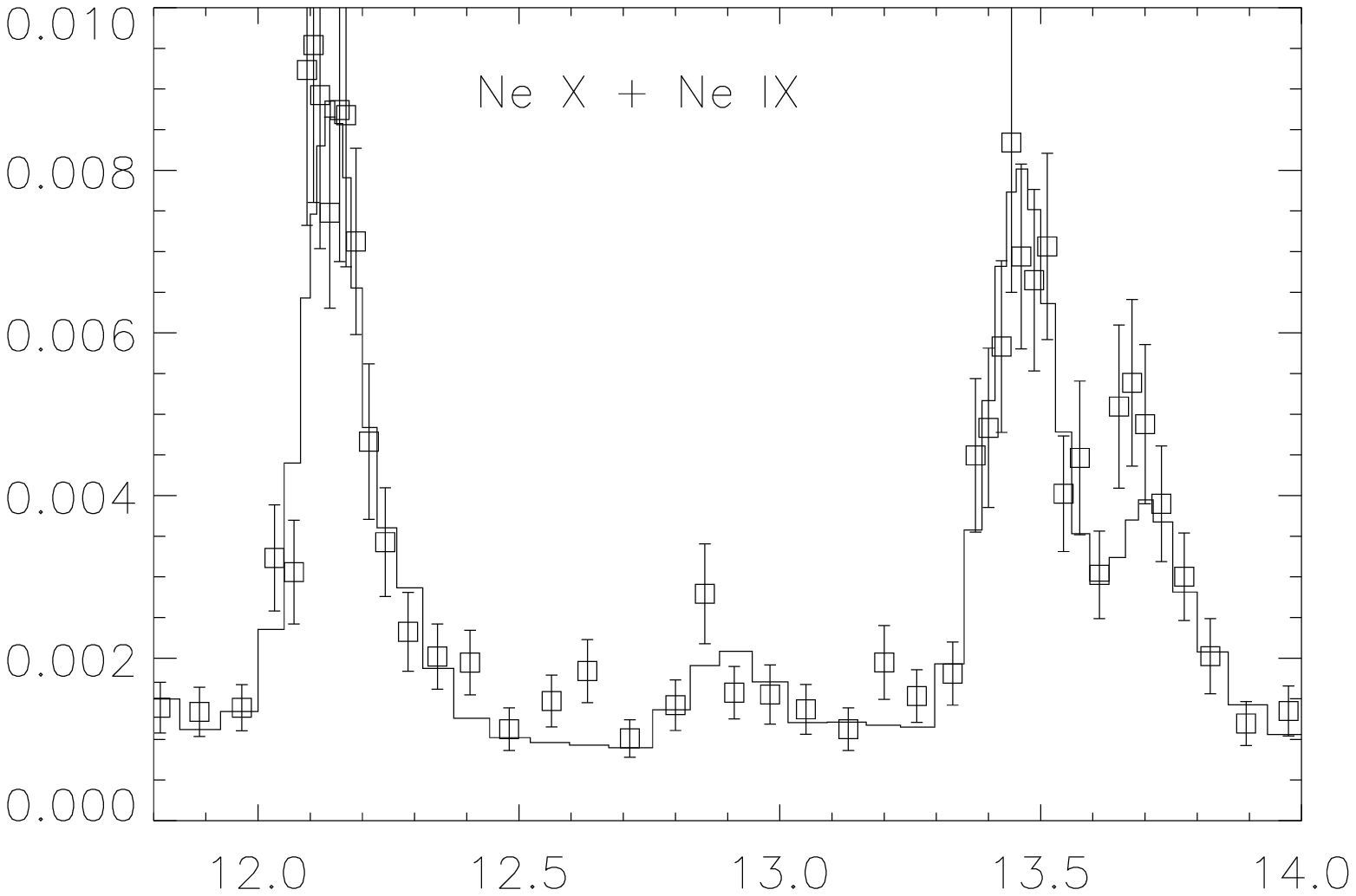}
\includegraphics[width=3in, height=1.85in]{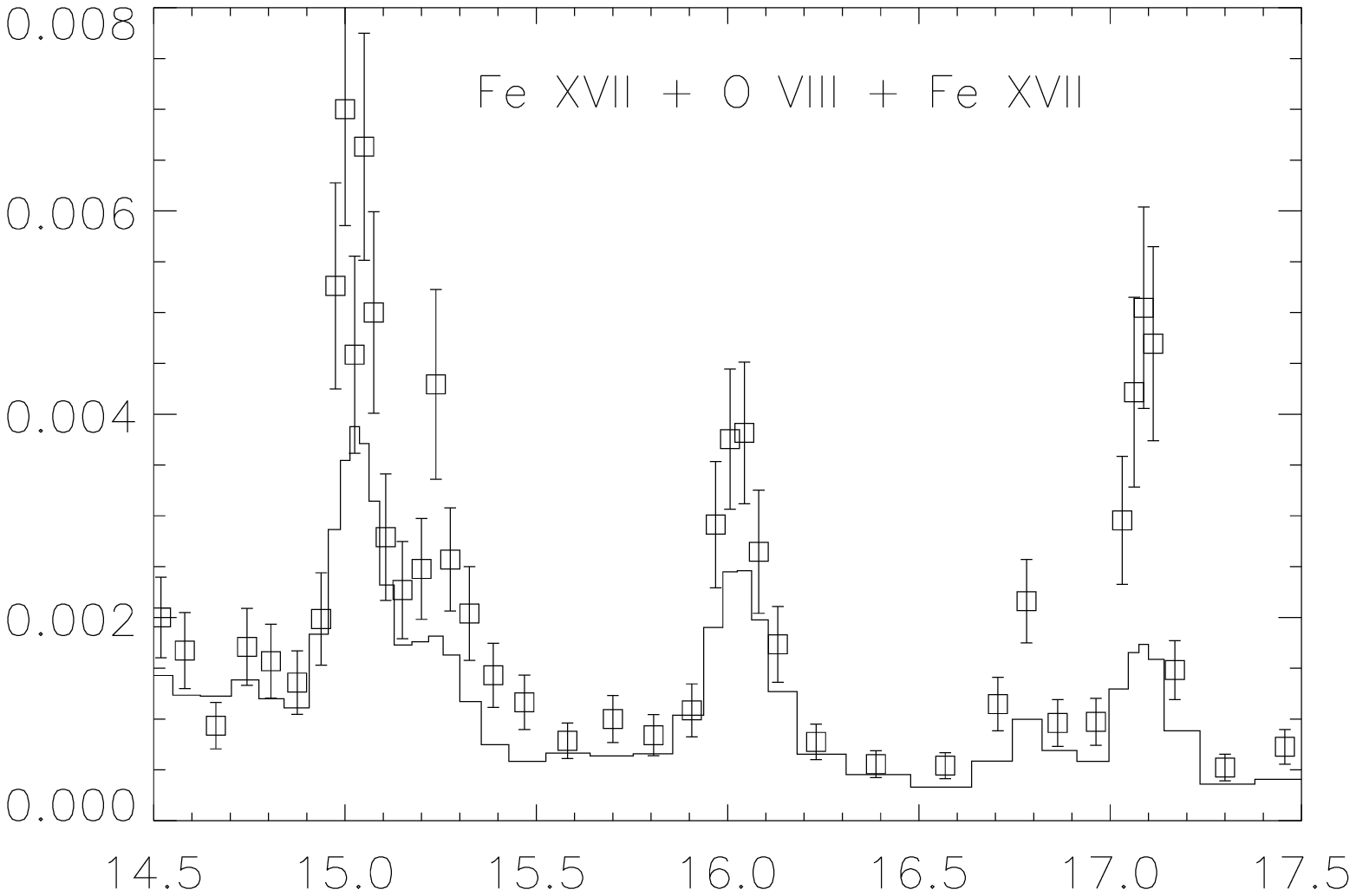}
\includegraphics[width=3in, height=1.85in]{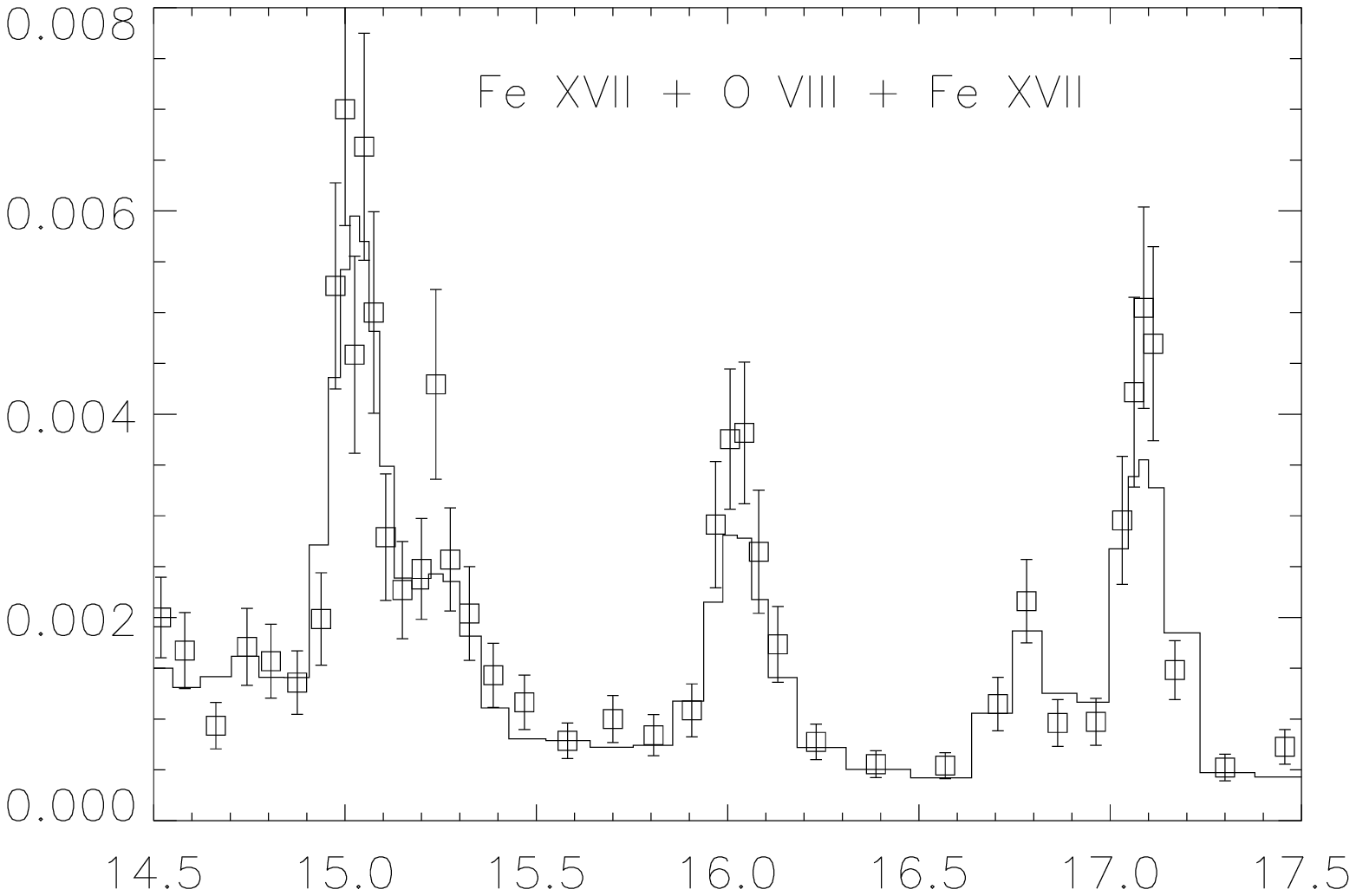}
\includegraphics[width=3in, height=1.85in]{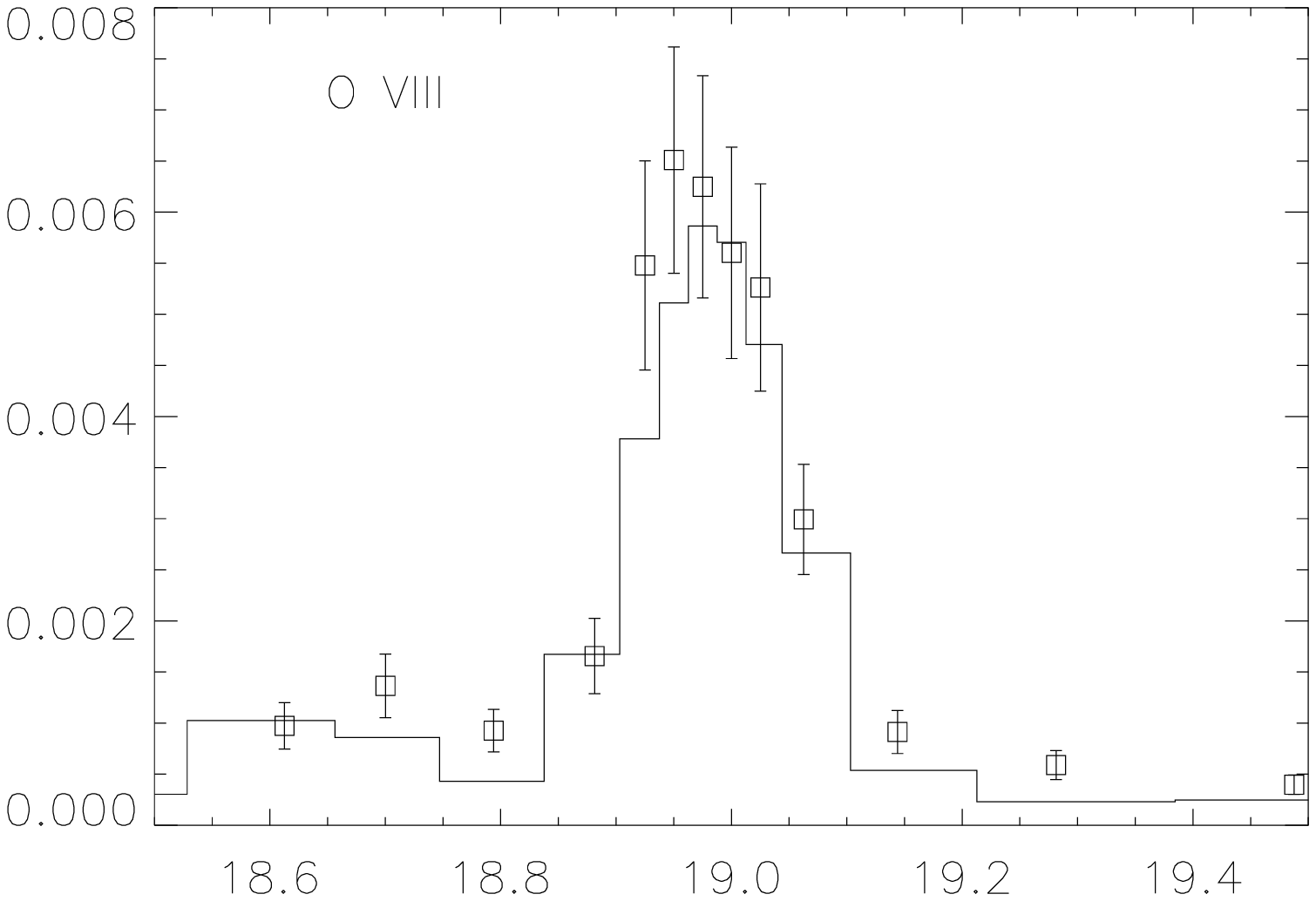}
\includegraphics[width=3in, height=1.85in]{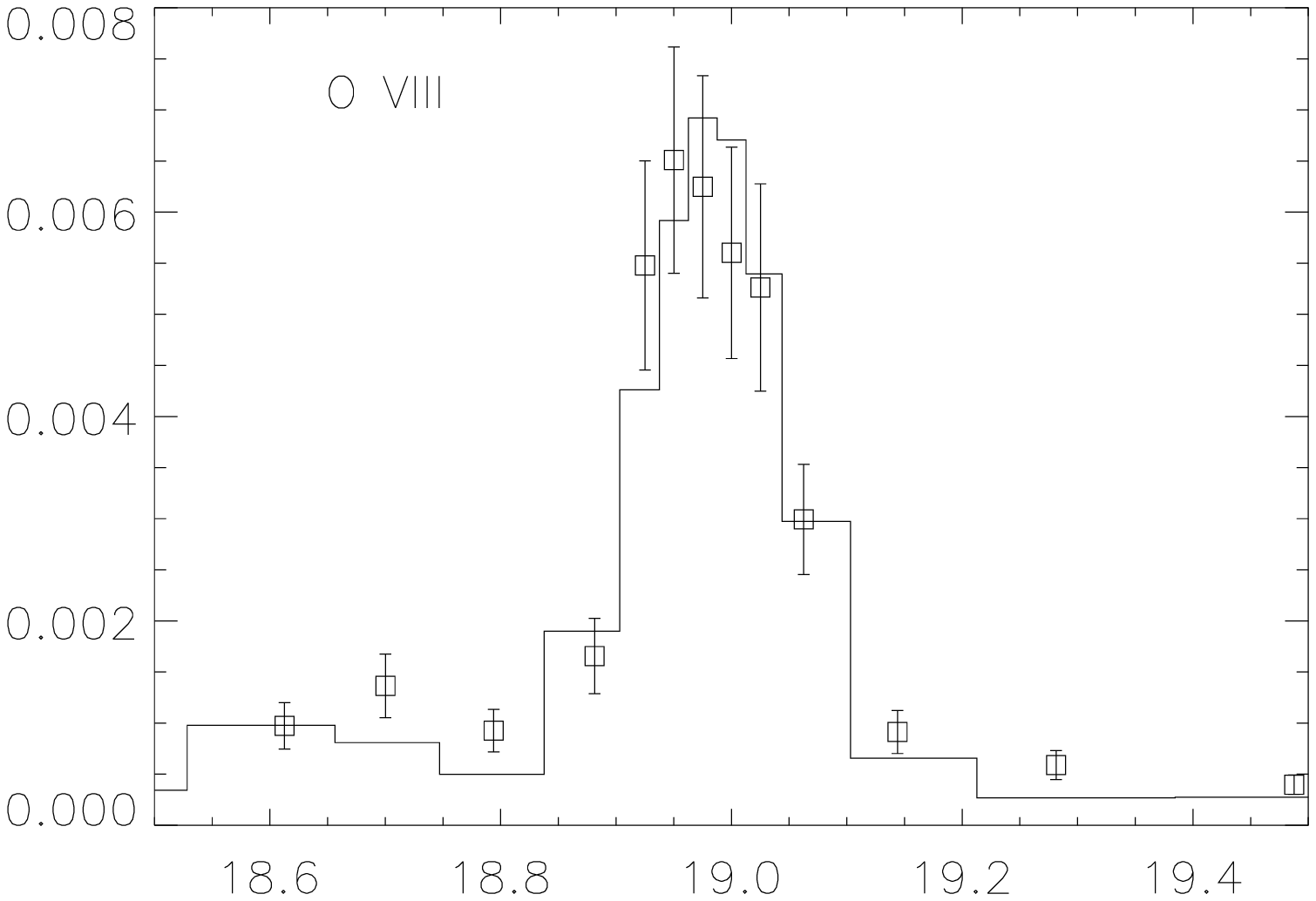}
\end{center}
\caption{ The positive 1-st order LETG X-ray
spectrum of SNR 1987A 
(rebinned to have a minimum of 30 cts per bin)
near some strong spectral lines :
(i) empty squares with $1\sigma$-error bars -- observed spectrum;
(ii) solid line --  fits from the XSPEC 1-shock (left column) and
2-shock (right column) models (see Table \ref{tab:fit}).
Horizontal axes -- observed wavelength
(\AA); vertical axes -- flux density (photons s$^{-1}$ \AA$^{-1}$).
}
\label{fig:xspec}
\end{figure}

We fitted one- and two-shock models to the LETG spectra of SNR~1987A
(rebinned to have a minimum of 30 counts per bin). The results from
our model fits are shown in Table~\ref{tab:fit}.  We find that a fit
with one-shock model is statistically unacceptable and that it yields
an unreasonably small X-ray absorption column density.  
(This value is from 3 to 4 times smaller
than that derived from the previous X-ray observations, Michael et al.
2002; Park et al. 2004; 
whose results were consistent with the neutral hydrogen column density
towards SNR 1987A deduced from observations in the optical and UV, 
Fitzpatrick \& Walborn 1990; Scuderi et al. 1996). 
Figure~\ref{fig:xspec} also illustrates 
how difficult is for the one-shock model to give a good fit to sectral
lines of ionic species which require different plasma conditions
(e.g., Si XIII and Si XIV with ionization potential, IP, of 2.67 and 2.44
keV and Fe XVII with IP of 1.27 keV).
Therefore, we
favor the two-shock model.  The results for the line broadening from
the global fit are close to those derived in Paper I: $z_0 =
1.13\times10^{-3}~[7.4\times10^{-4}-1.54\times10^{-3}]$~vs.
$1.57\times10^{-3}$; $\lambda_0 = 0.061~[0.057-0.064]$~vs. 0.047~\AA;
and $\alpha = -1.17~[-1.94 - -0.08]$~vs. -1.3 (the 90\%-confidence
limits are given in brackets). The relatively small differences can be
attributed to the fact that the global fitting procedure
simultaneously handles both lines and continuum.  

As a check, we also fitted one- and two-shock models to the
pulse-height spectrum of the zero-order LETG image. The best-fit
(two-shock) results are shown in Table~\ref{tab:fit}.  Of course, the
parameters determined from the pulse height spectrum, especially the
abundances, are less tightly constrained than those determined from
the dispersed images.  However, agreement of the fit to the
pulse-height spectrum with the fit to the dispersed spectrum gives us
confidence in using also the pulse-height spectrum to infer element
abundances and conditions in the X-ray emitting gas.  

The abundances of SNR~1987A derived here are consistent with those
derived from previous CCD spectra (Michael et al. 2002, Park et al.
2004).  We find that the abundances of Ne, Mg, Si, S and Fe are very
close to those found in X-rays for the LMC SNRs (Hughes et al. 1998).
On the other hand, the N and O abundances derived
here are lower by a factor of about two than those found 
for the inner ring by Lundqvist \& Fransson (1996).  
We note that similar lower  nitrogen abundance
was found by Pun et al. (2002) in their analysis of
the optical and UV spectrum of Spot 1 on the inner circumstellar ring.  
But, it is interesting to note that the relative abundance of nitrogen
and oxigen derived in our analysis, N/O $= 1.1$ (by number), is consistent
with that derived from the modeling of the optical/UV emission
of the inner ring (Lundqvist \& Fransson 1996, Sonneborn et al. 1997).
Thus, we confirm the nitrogen enhancement in the CSM around
SNR~1987A, which strongly suggests that the CSM consists of matter 
that had undergone CNO processing in progenitor star and was 
subsequently ejected.  

Significant departures from electron-ion temperature equilibration may
be present in the post-shock plasma (Michael et al 2002).  In the fits
we present here, we have not included this effect.  We have explored
models including departures from electron-ion temperature
equilibration (the {\it vnpshock} option in XSPEC) and we find that
they give equally good fits to the spectra, but a detailed analysis of
such models is beyond the scope of this paper (see also \S
\ref{sec:disc}).

\subsection{Distribution of Shocks} 
\label{subsec:DS} 
One of the basic
results from the line-profile analysis of Paper I was that a
distribution of shocks with velocities in the range $340-1700$\kms is
compatible with the LETG data. Given the relation between the
postshock temperature and the shock velocity for strong adiabatic
shocks with electron-ion temperature equilibration, 
\begin{equation}
kT_e = \frac{3}{16}\mu V_S^2 = 1.4 [V_S/1000~\mbox{km
s}^{-1}]^2\hspace{0.5cm}\ \mbox{keV;} 
\label{eqn:Tshock}
\end{equation} 
where $\mu = 0.72 m_p$~for SNR~1987A.  The observed velocity
range corresponds to a temperature range of $kT_e = 0.15 - 4$~keV.

To develop this scenario, we constructed a new XSPEC model having the
following features: (1) the distribution of emission measures in the
shocked gas is determined from the Chebyshev polynomial algorithm as
is used in the standard XSPEC $c6pvmkl$ model (Lemen et al. 1989); (2)
the basis vectors for X-ray emission from the distribution of shocks
are those from the XSPEC $vpshock$ model; (3) all shocks share the
same element abundances and X-ray absorption.  Moreover, to
decrease the number of free parameters, we also assumed: 
(4) the ionization age
$(n_et)_i$ of each individual shock $i$ is a power-law function of the
postshock temperature, {\it i.e.,} $(n_et)_i \propto (kT_e)^p_i$~where 
$p$ is a free parameter. Details about the line broadening are given below.

As in our earlier analysis, we assume the emission lines to be Gaussian
profiles with FWHM depending on the wavelength. We recall that the
wavelength dependence of the line width represents the physical notion
that faster shocks produce higher-temperature plasma whose emission is
dominated by lines at shorter wavelengths. Now, by fitting the entire
spectrum with models having a distribution of shock velocities, we can
test this assumption in a more rigorous way.  Namely, we assume that
the {\it total} X-ray spectrum of a parcel of shocked plasma (having
temperature $T_e$) is moving with postshock velocity given by equation
(\ref{eqn:Tshock}).  Then, similarly to eq.~(\ref{eqn:stratified})
we write: \begin{equation} \Delta \lambda_{tot} = 2 \Delta \lambda_0
\pm 2 z_S \lambda\,,~ \label{eqn:stratified1} \end{equation} where
$z_S=\beta (3/4)V_S/c$~;~$V_S$ is the shock velocity; (3/4) stands for
the case of a strong shock entering a stationary gas; and $c$ is the
speed of light. The parameter $\beta$ ($0 \leq  \beta \leq 1$) allows
for various geometrical effects such as viewing angle to be considered
as well.

With these assumptions, we can find a satisfactory ($\chi^2/dof =
414/431$) simultaneous fit to the positive and negative LETG spectra
of SNR~1987A.  Figure \ref{fig:DS} shows the distribution function of
emission measures and ionization ages determined by the fit.  Note
that the shape of the inferred shock distribution is {\it bimodal}
peaking very close to the temperature values derived in our discrete
two-shock model fit. To verify this conclusion, we tried to fit
various models having more X-ray plasma with temperature between the
two peaks.  In no case could we find an acceptable fit without a
bimodal distribution.  Note also that the ionization age of the
shocked gas decreases with increasing shock temperature.  

The absorption column density and element abundances have the values
(given are $1\sigma$ errors): $N_H
= 1.52\pm0.02\times10^{21}$~cm$^{-2}$, N$=0.88\pm0.13$,
O$=0.10\pm0.01$, Ne$=0.31\pm0.02$, Mg$=0.25\pm0.02$, Si$=0.29\pm0.02$,
S$=0.48\pm0.09$, and Fe$=0.16\pm0.02$. We note that these parameters
are also very close to those determined from the two-shock model fit
(for comparison see Table \ref{tab:fit}).
Moreover, the derived value of $\beta=0.35\pm0.06$ is close to that
expected for a radial gas motion in a disk viewed at inclination angle
$i$: $\overline{\sin\phi} \sin i = 0.45$ ($ \overline{\sin\phi} =
2/\pi~, 0\le\phi\le\pi/2$; $i = 45^{\circ}$ for the inner ring;
Sugerman et al. 2002).  These results give us confidence in our global
fits to the LETG spectrum of SNR~1987A.

\section{MARX simulations} 
\label{sec:marx} 
As a final test of our
model, we used the MARX\,\footnote{See http://space.mit.edu/CXC/MARX/}
software to simulate the actual two-dimensional images and dispersed
spectra which incorporate the kinematic and spectral information
deduced from our analysis of the line profiles, line ratios and global
fits to the observed X-ray spectra of SNR~1987A.  Since the object is
spatially resolved by Chandra, our simulations take into account the
actual brightness distribution of the X-ray emission as well.  The
simulations correspond to the actual observing conditions (e.g. date
of observation, exposure time, pointing of the telescope, roll angle
etc.).

\begin{figure} 
\begin{center} 
\includegraphics[width=4.16in, height=3in]{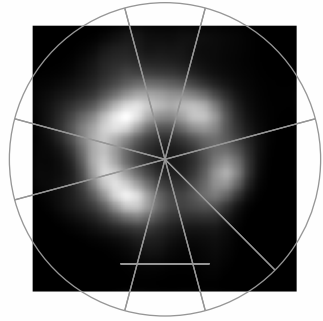} 
\end{center} 
\caption{ The deconvolved 0-th order
image from the LETG observations of SNR~1987A. North is up and East is
on the left.  Shown are the boundaries of the eight sectors
(subimages) used in the MARX simulations (\S \ref{sec:marx}). The
horizontal line (1 arcsec) indicates the scale of the image.  }
\label{fig:marx0} \end{figure}

The actual brightness distribution of the X-ray emission used in our
simulations is shown in Fig. \ref{fig:marx0}. This is the 0-th order
image from our LETG observations deconvolved according to the same
procedures used in our imaging observations (Burrows et al. 2000; Park
et al. 2002, 2004). Shown also are the eight sectors we used to
represent the details of the image morphology.

To construct the input spectra, we assume that the X-ray emitting
plasma is confined in the plane of the inner ring and that it is
expanding radially.  
Locally, we assume a distribution of
shocks with characteristics as described in \S \ref{subsec:DS} (see
also Fig. \ref{fig:DS}), with emission measure scaled to correspond
to the brightness distribution in each sector.  

We constructed our MARX simulations according to the following
procedure.  For each sector, we simulated the observed spectrum,
assuming that the X-ray spectrum corresponds to the physical picture
given above and a flux scaled to the total observed flux from
SNR~1987A according to the fractional brightness of that sector in the
zero order image. We determined the red- or blue-shift of the spectrum
of each sector according to its postshock plasma velocity, corrected
for its position (azimuthal angle), and the inclination of
the inner ring ($i = 45^{\circ}$).  
We then concatenated the
resultant simulated data from the all sectors to derive a data set for
the entire object.  We then extracted the simulated LETG spectra
following the CIAO Science Threads for Grating Spectroscopy.  In order
to minimize the statistical fluctuations in the simulated spectra, we
averaged the ten different realizations of the simulated spectra. 

\begin{figure} 
\begin{center} 
\includegraphics[width=3in, height=1.85in]{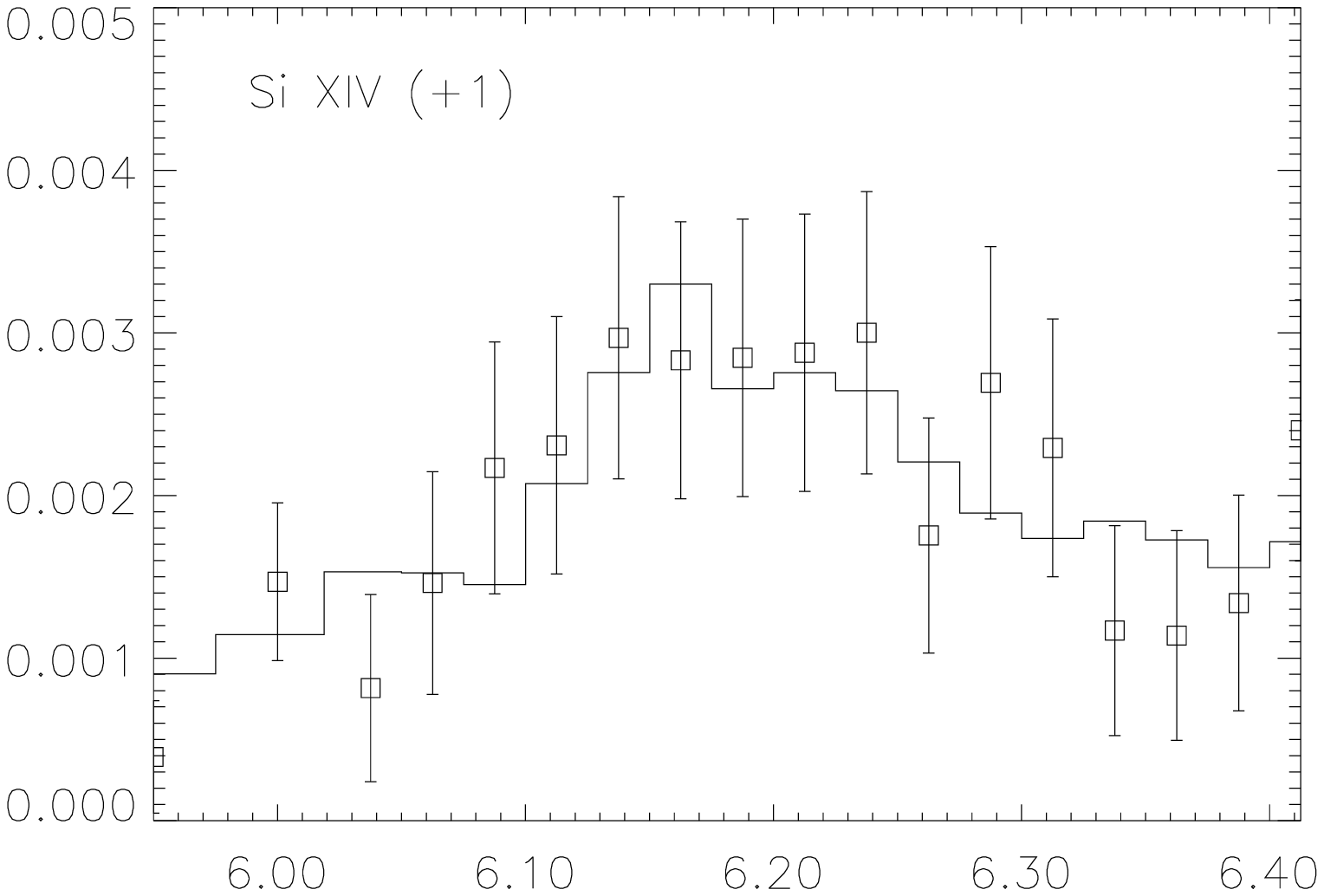} 
\includegraphics[width=3in, height=1.85in]{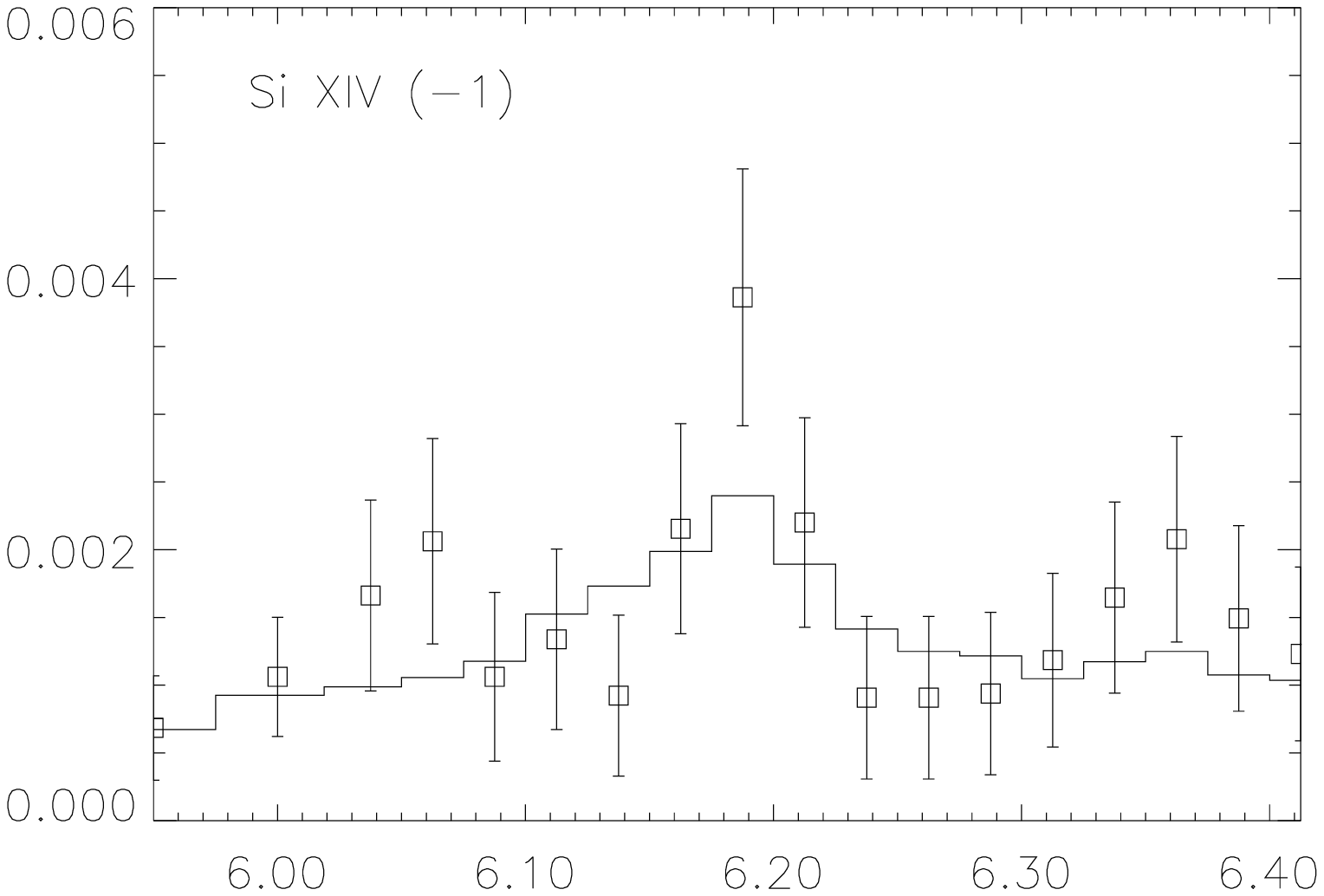} 
\includegraphics[width=3in, height=1.85in]{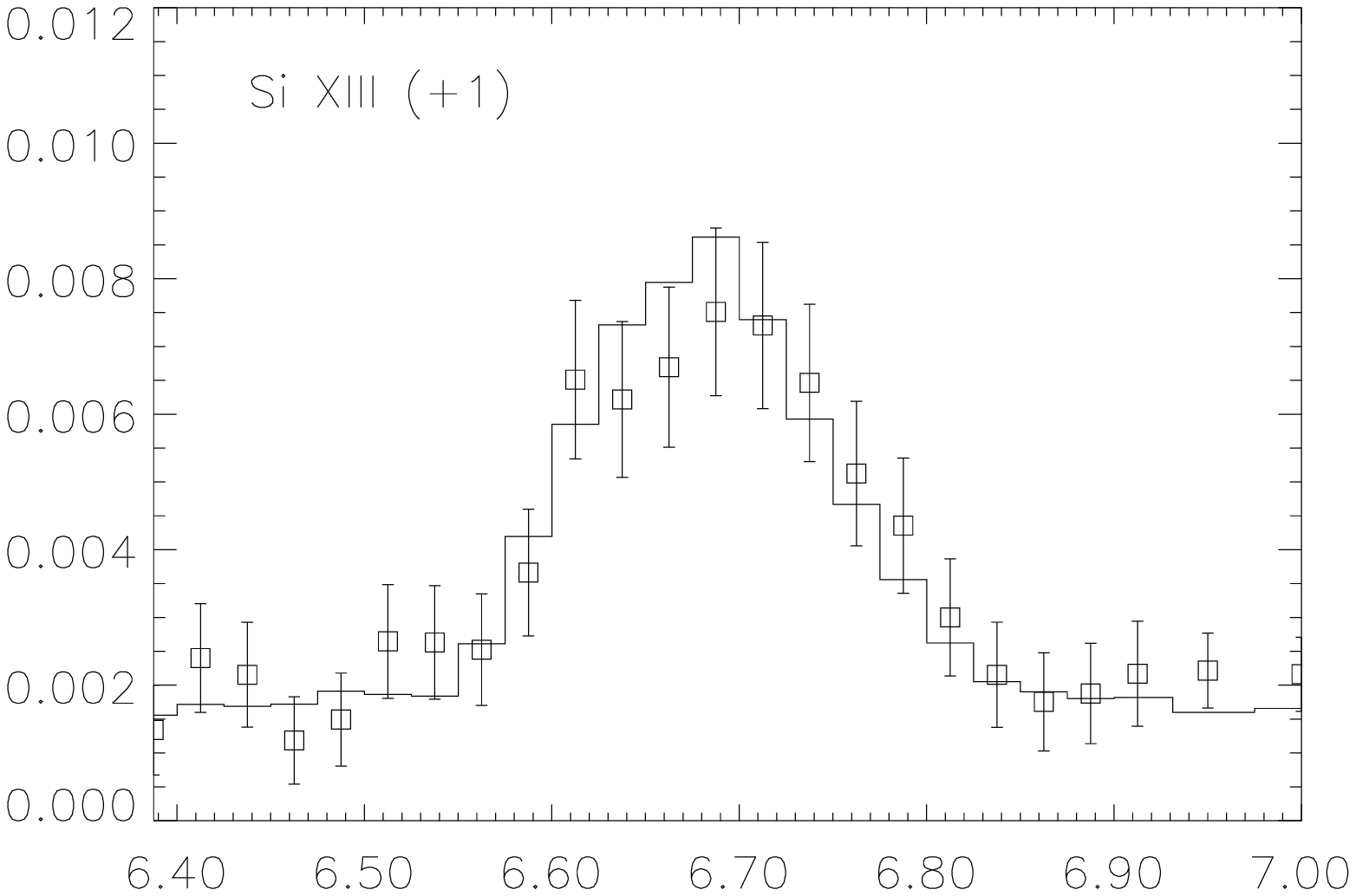} 
\includegraphics[width=3in, height=1.85in]{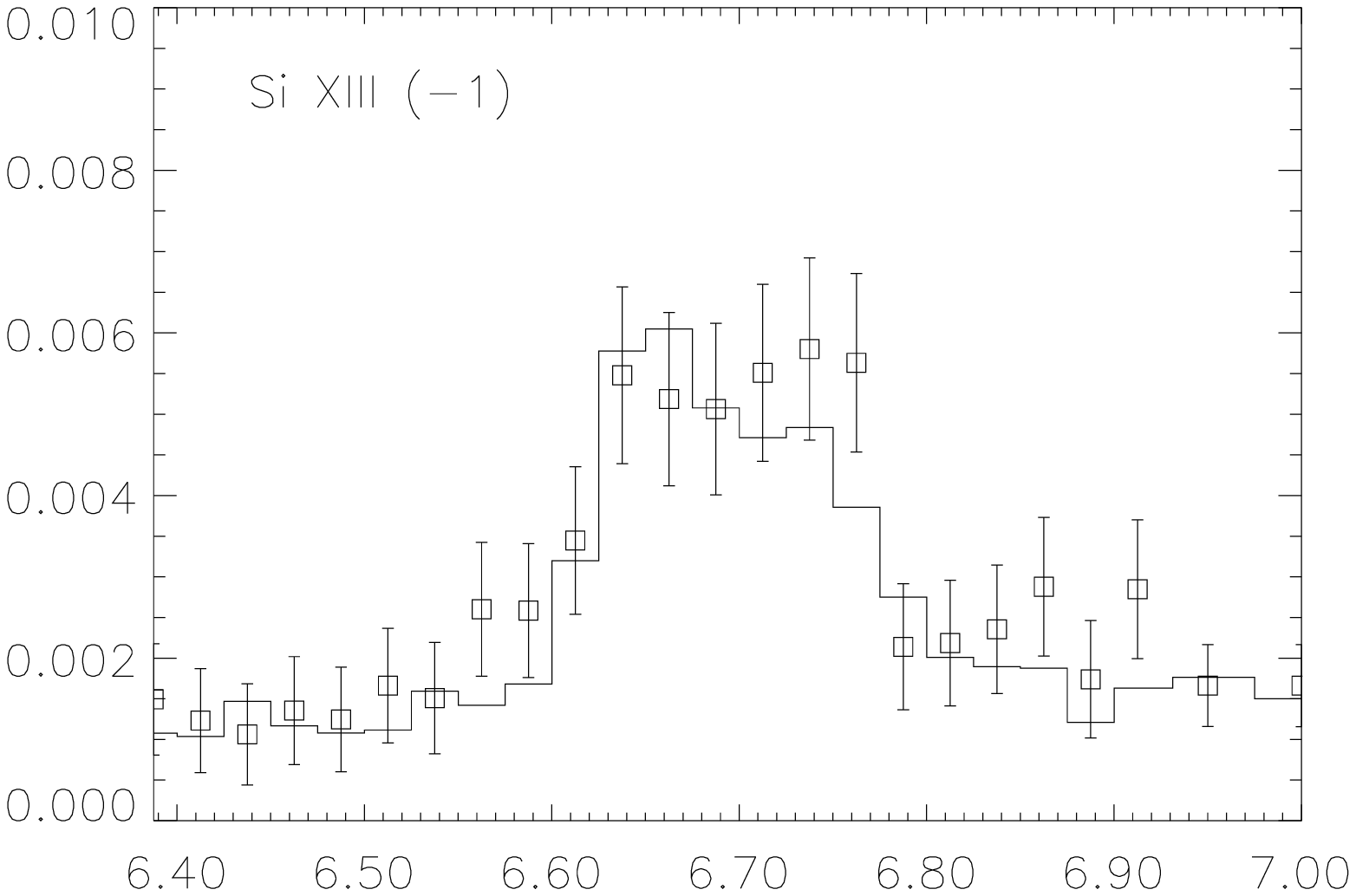} 
\includegraphics[width=3in, height=1.85in]{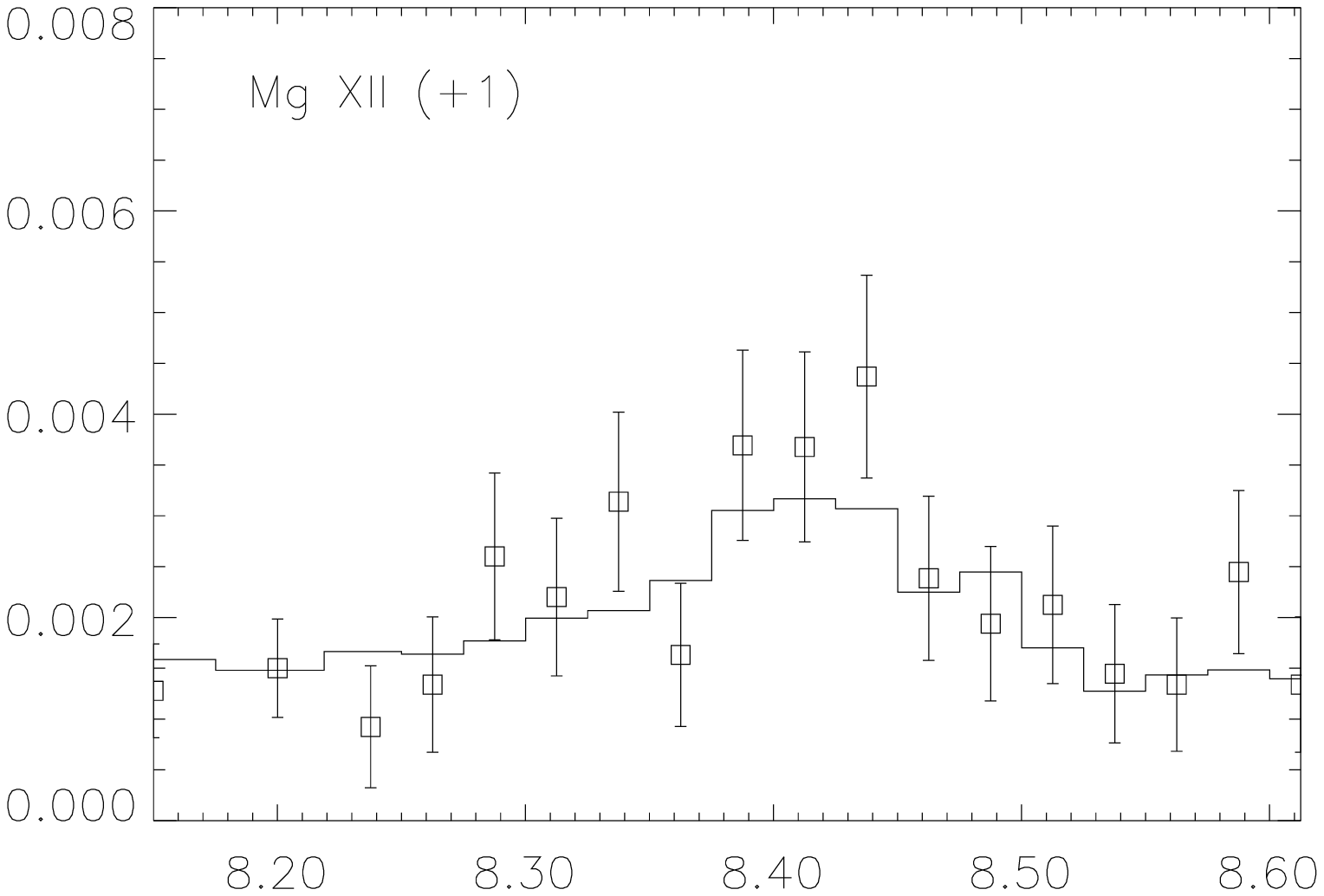} 
\includegraphics[width=3in, height=1.85in]{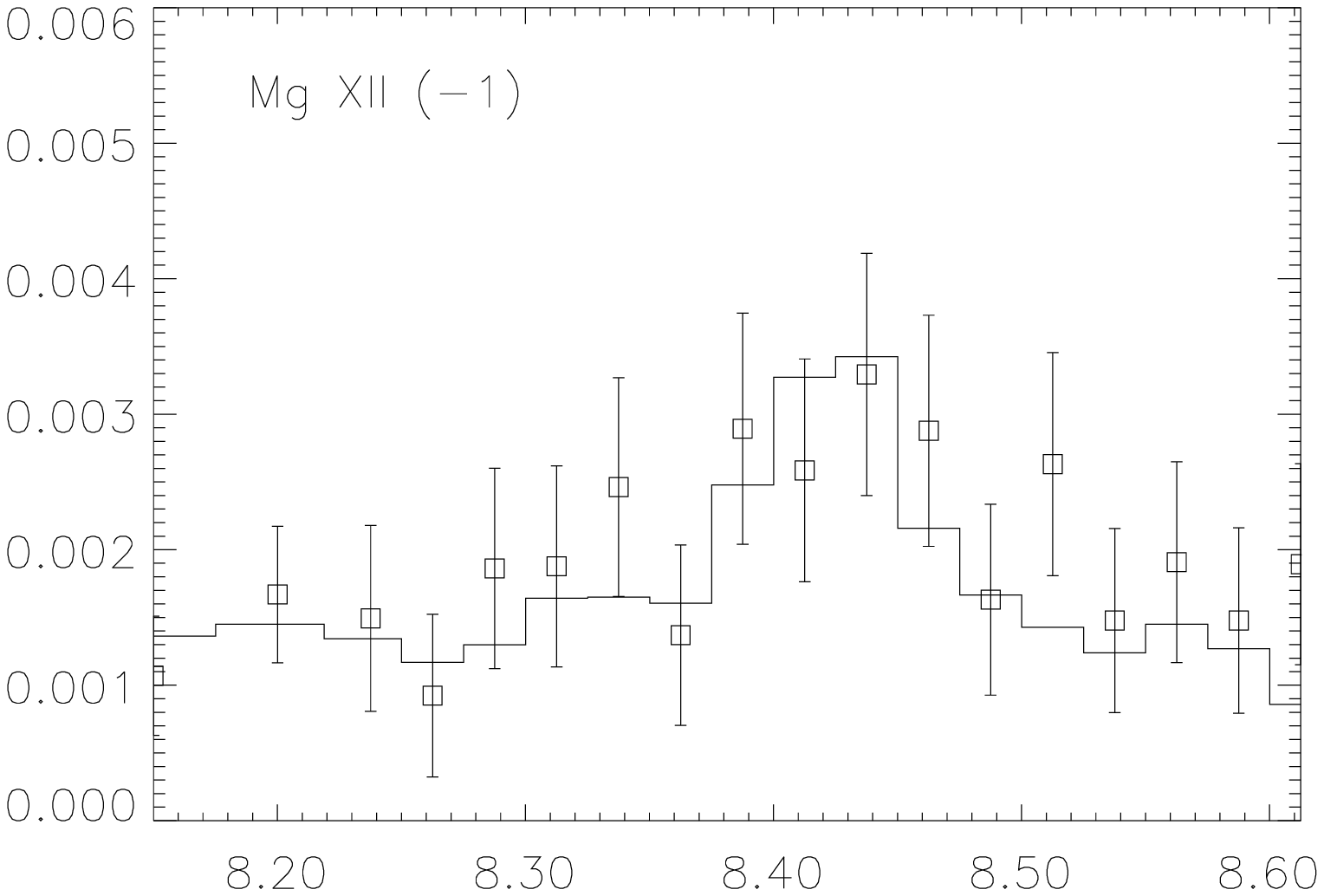} 
\includegraphics[width=3in, height=1.85in]{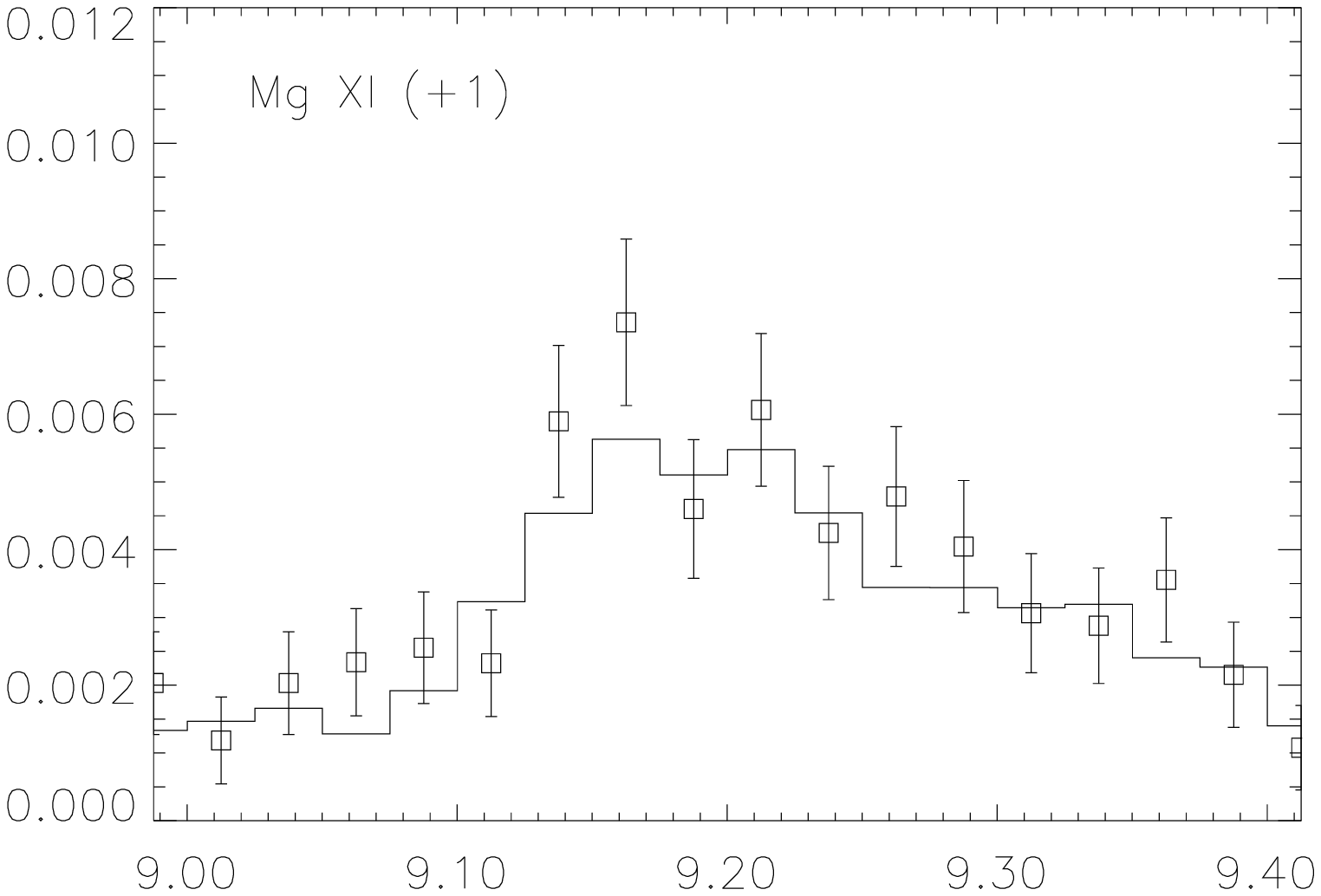} 
\includegraphics[width=3in, height=1.85in]{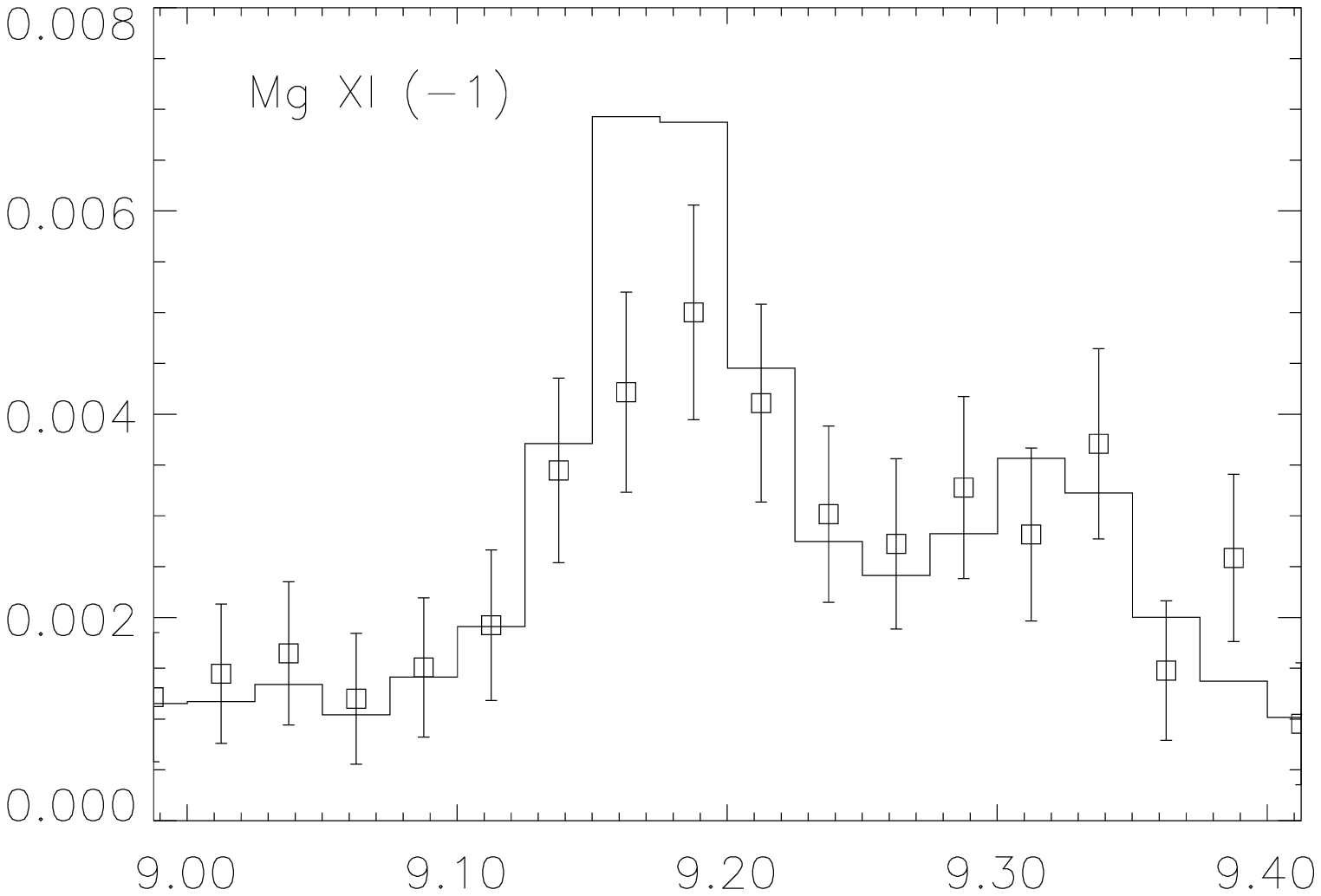} 
\end{center} 
\caption{ The X-ray
spectrum of SNR 1987A near the strong spectral lines of the H-like and
He-like ions Si XIV, Si XIII and Mg XII, Mg XI: (i) empty squares with
$1\sigma$-error bars -- observed spectrum; (ii) solid line --
simulated MARX spectrum.  Horizontal axes -- observed wavelength
(\AA); vertical axes -- flux density (photons s$^{-1}$ \AA$^{-1}$).
The positive and negative 1-st order LETG spectra are shown in the
left and right columns, respectively.  
} 
\label{fig:marx1}
\end{figure}

\begin{figure} 
\begin{center} 
\includegraphics[width=3in, height=1.85in]{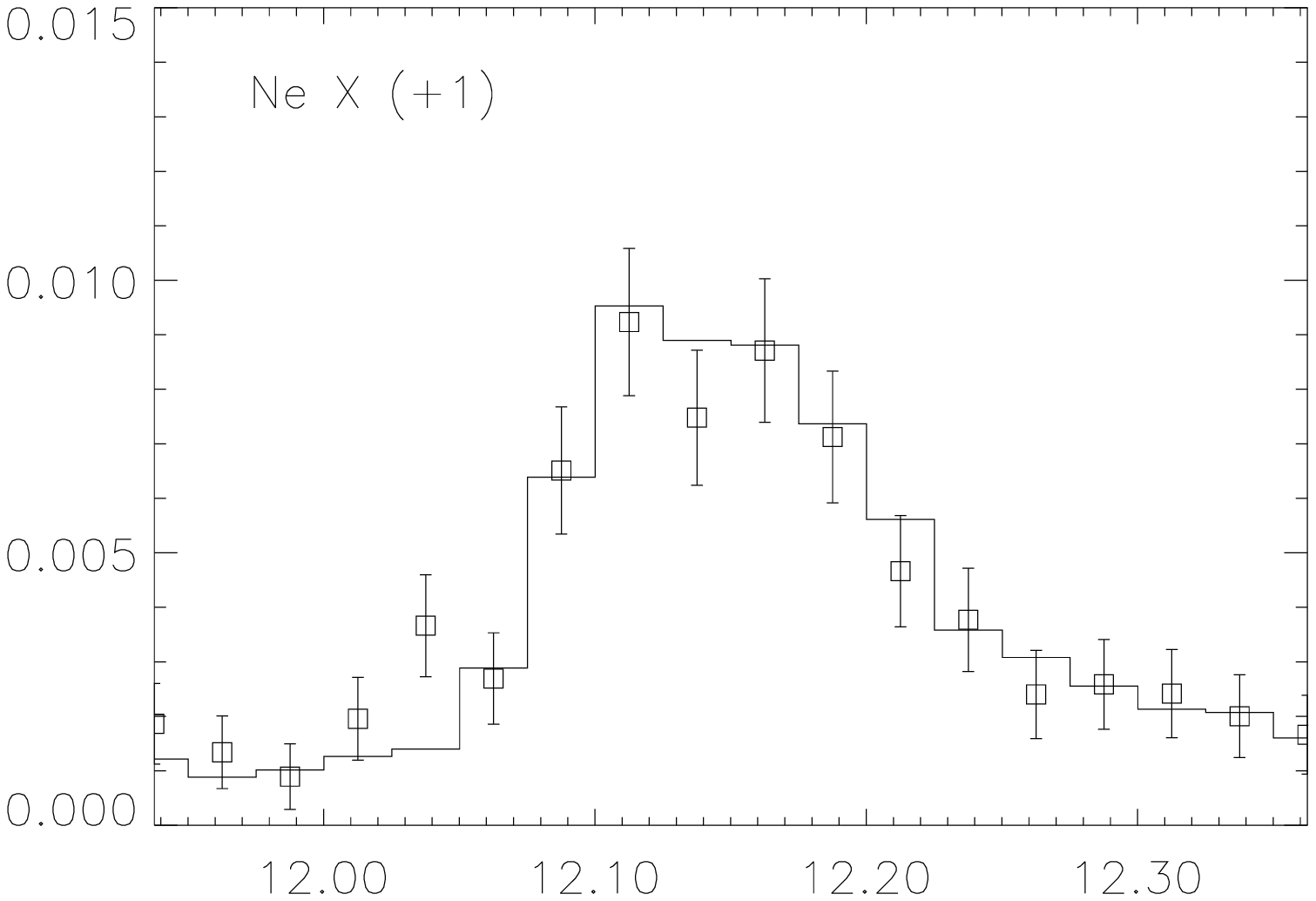} 
\includegraphics[width=3in, height=1.85in]{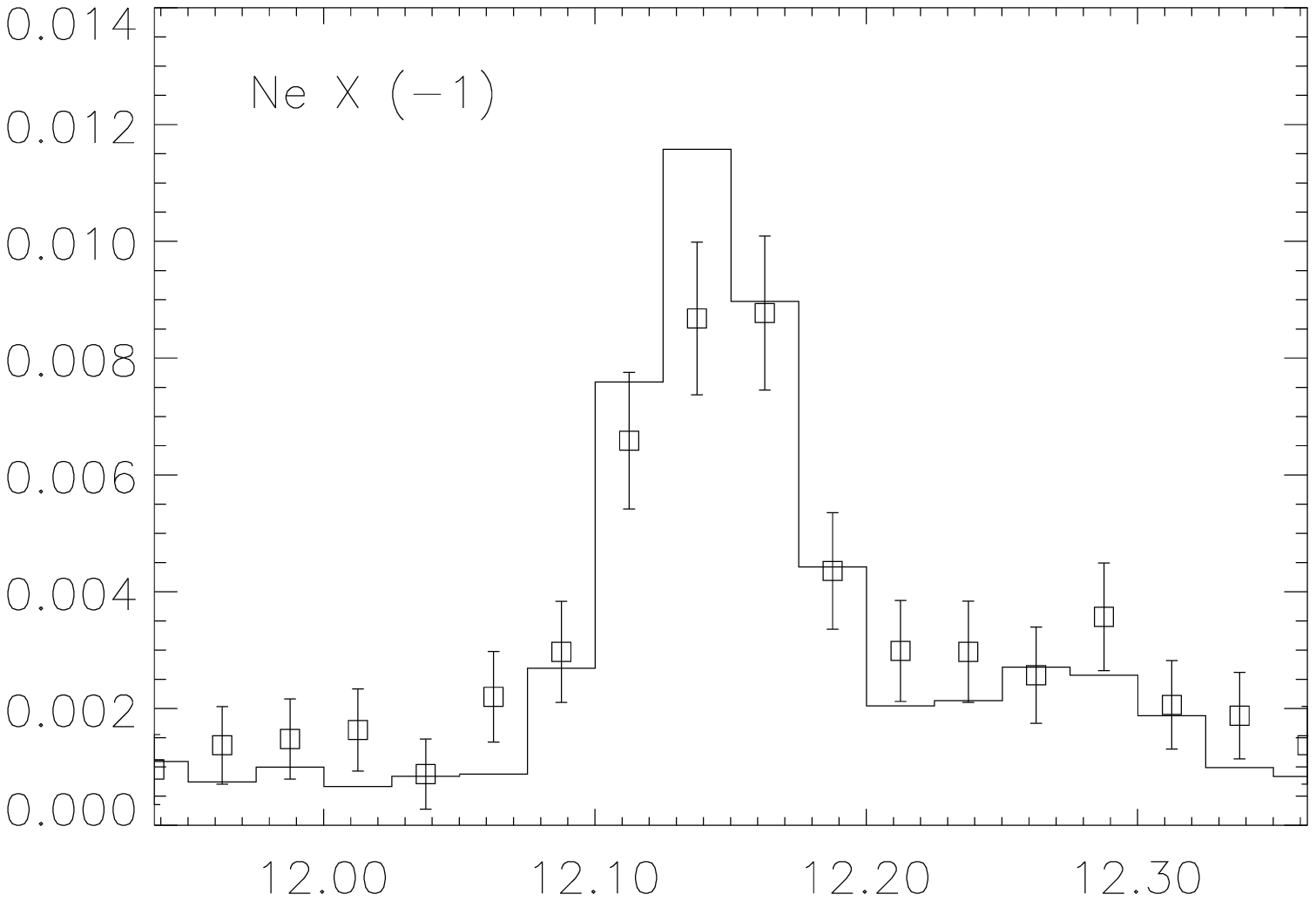} 
\includegraphics[width=3in, height=1.85in]{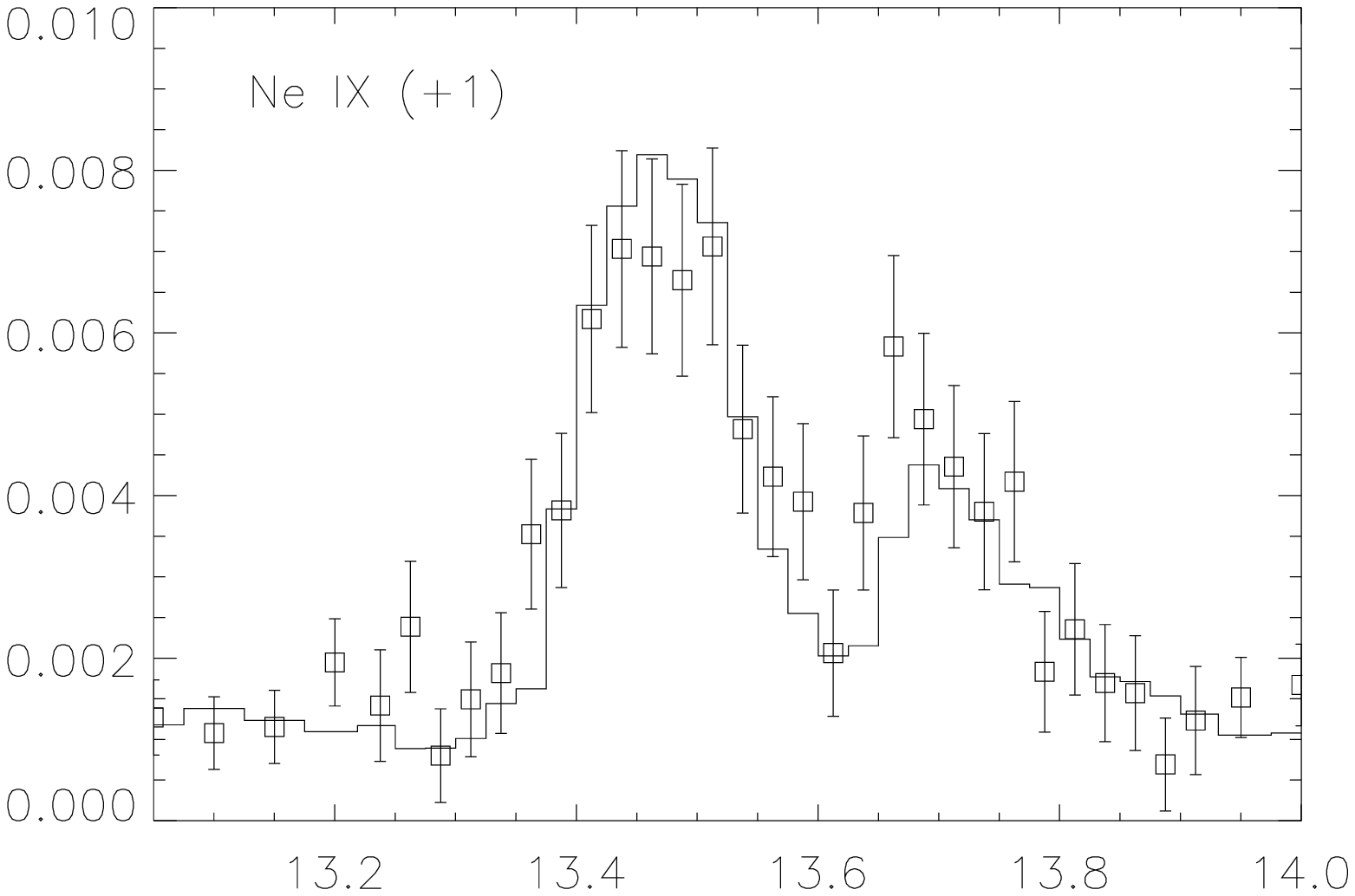} 
\includegraphics[width=3in, height=1.85in]{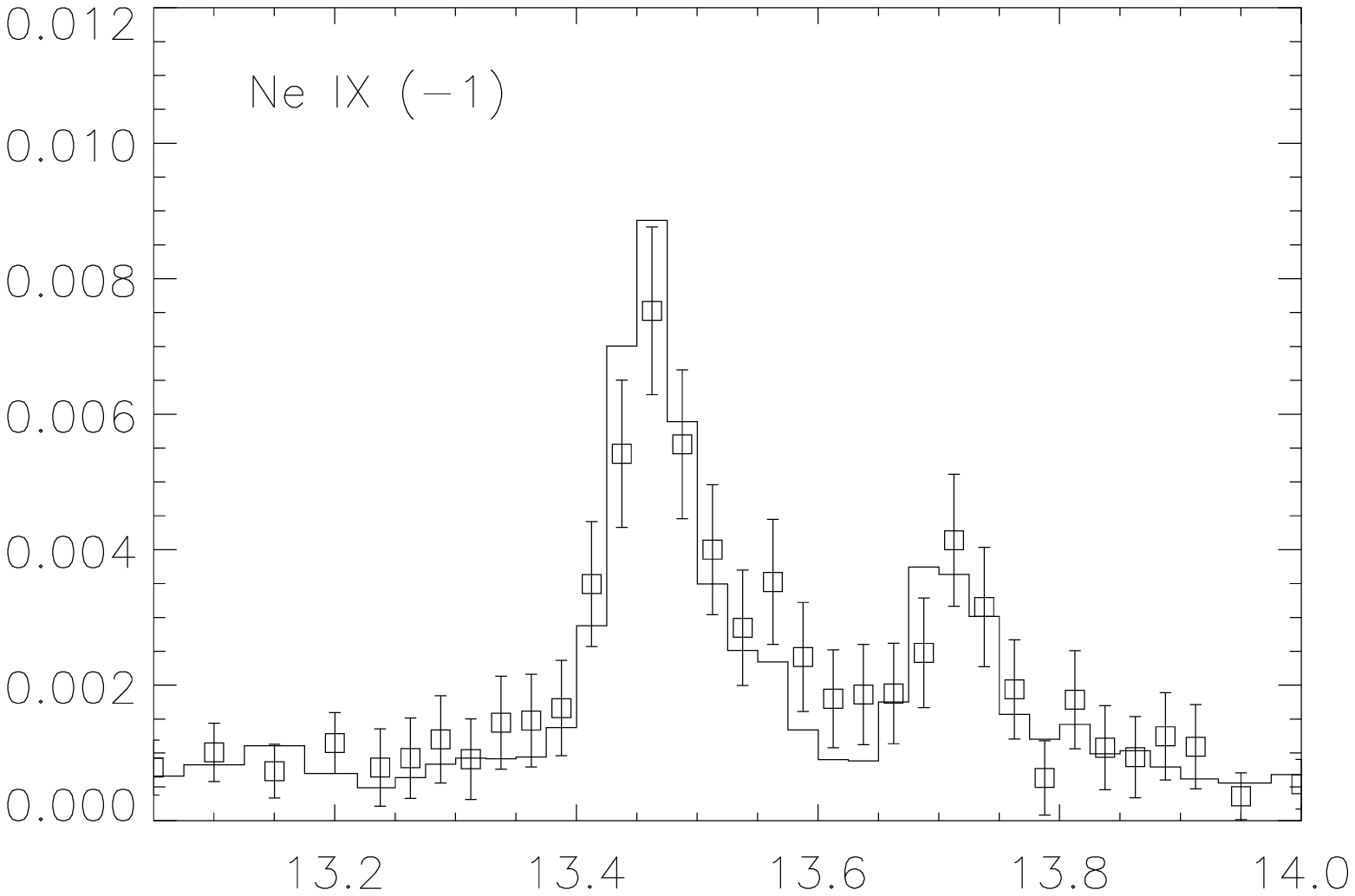} 
\includegraphics[width=3in, height=1.85in]{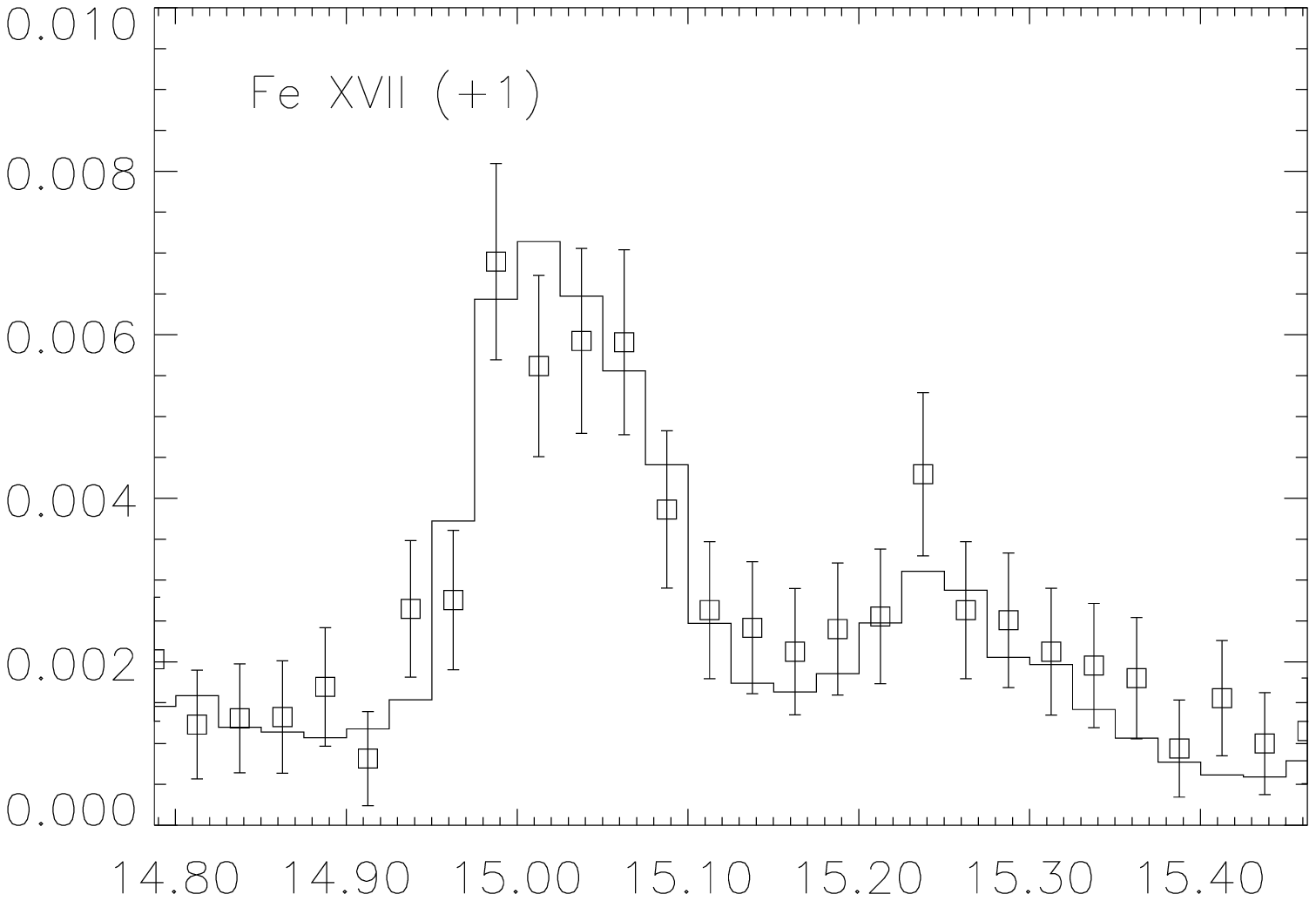} 
\includegraphics[width=3in, height=1.85in]{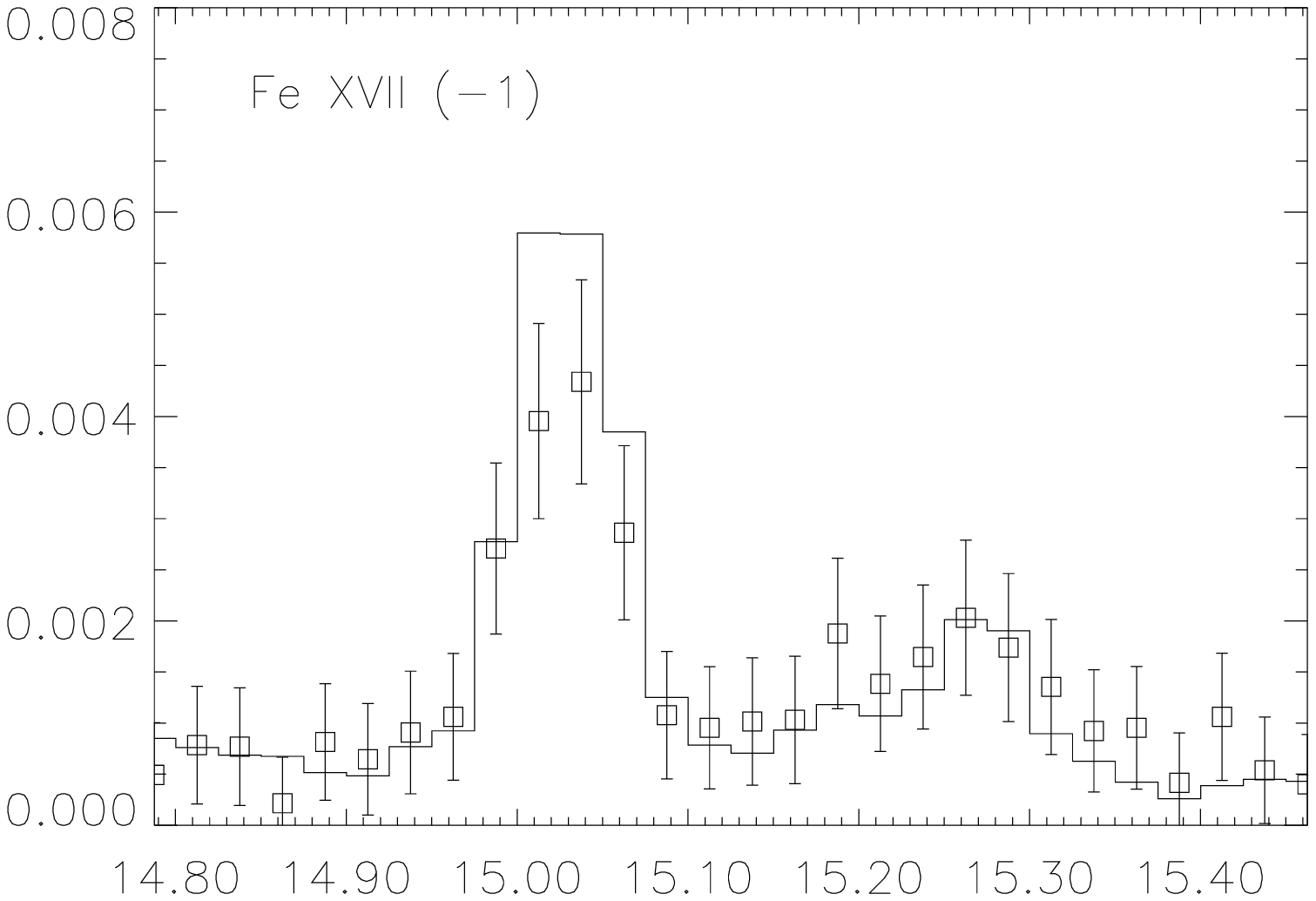} 
\includegraphics[width=3in, height=1.85in]{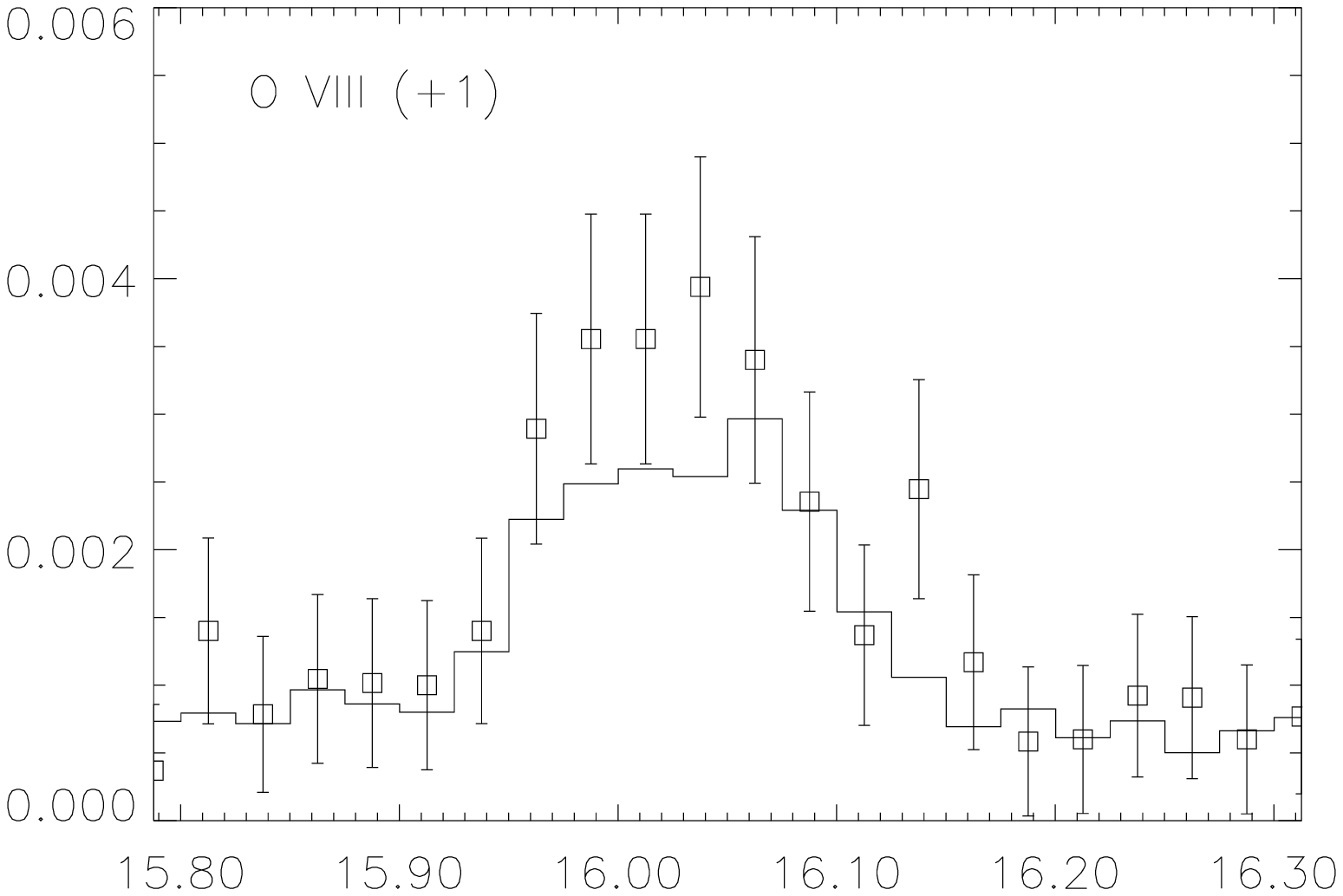} 
\includegraphics[width=3in, height=1.85in]{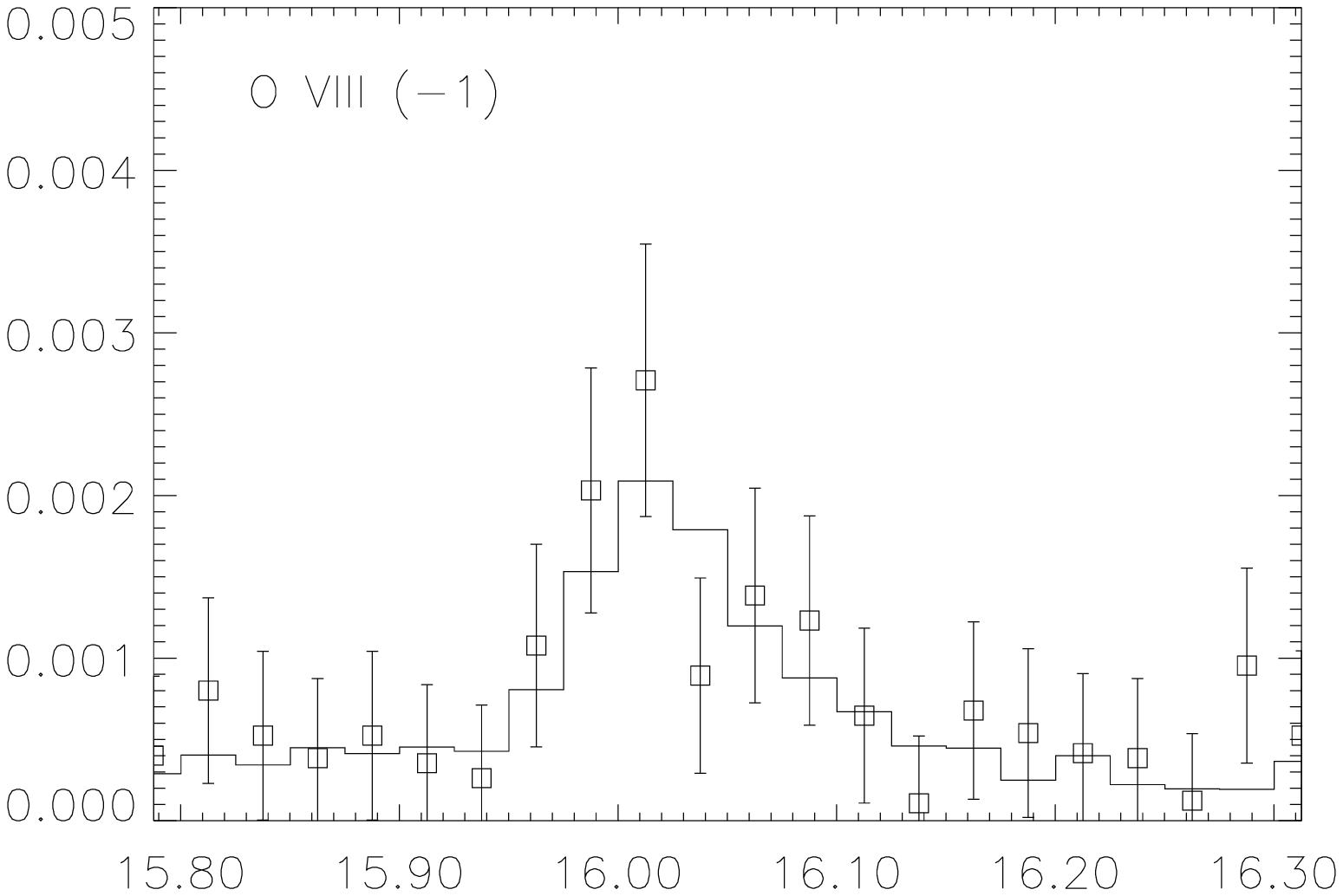} 
\end{center} 
\caption{ The same as
in Fig.~\ref{fig:marx1} but for the H-like and He-like ions iof Ne X
and Ne IX and the strong lines of Fe XVII and O VIII (Ly$_{\beta}$).
} 
\label{fig:marx2} 
\end{figure}

\begin{figure} 
\begin{center} 
\includegraphics[width=3in, height=1.85in]{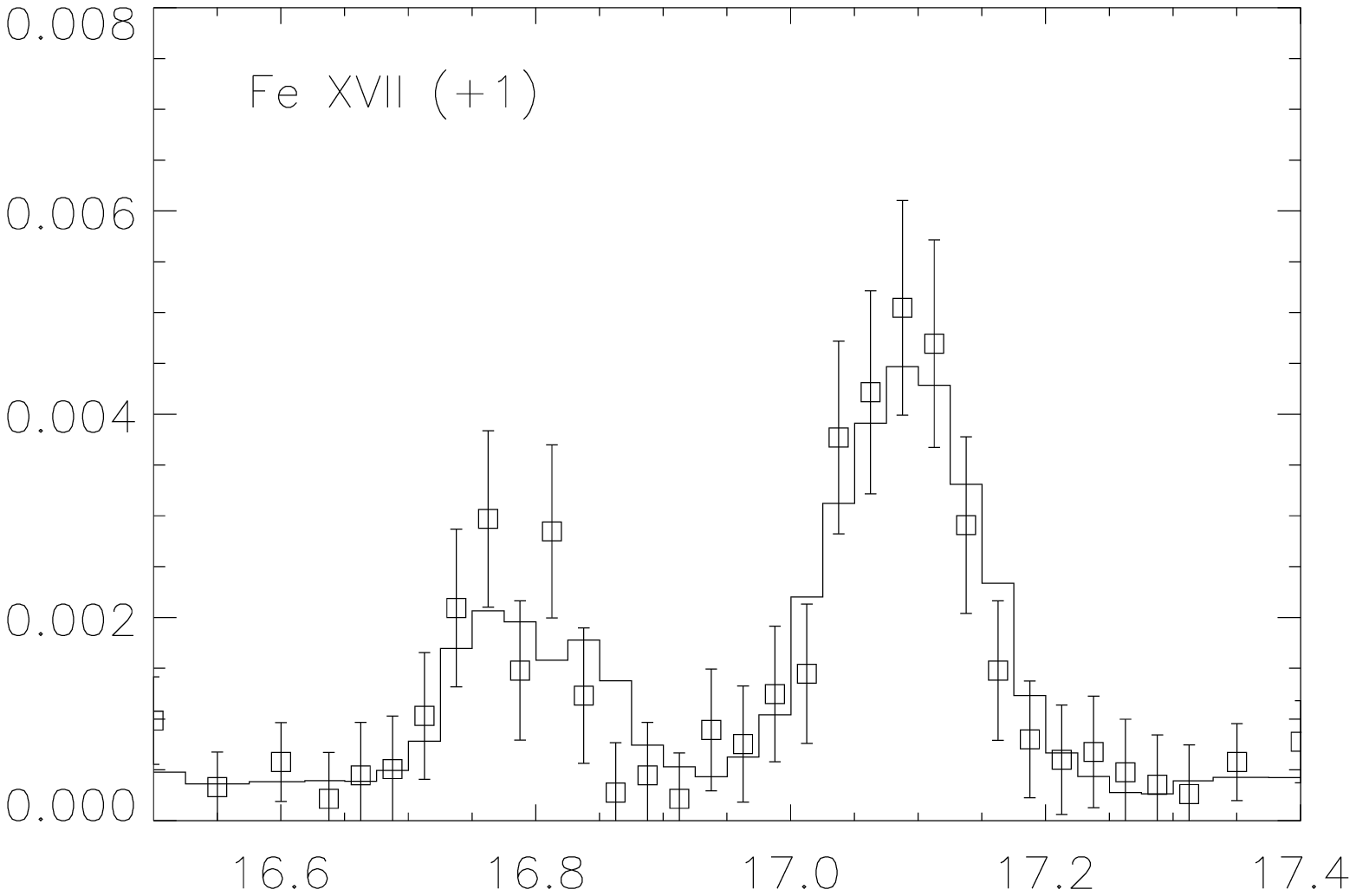} 
\includegraphics[width=3in, height=1.85in]{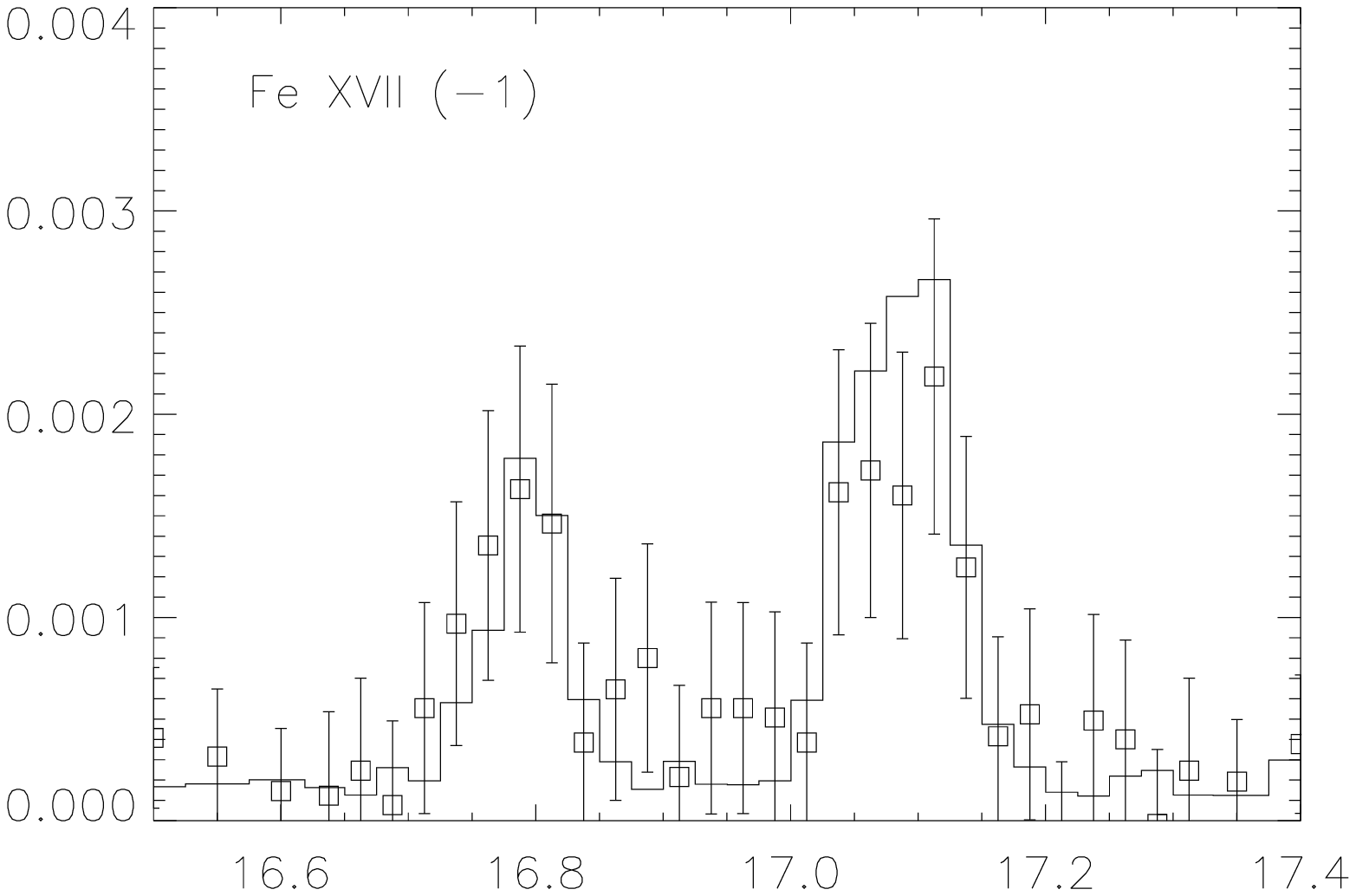} 
\includegraphics[width=3in, height=1.85in]{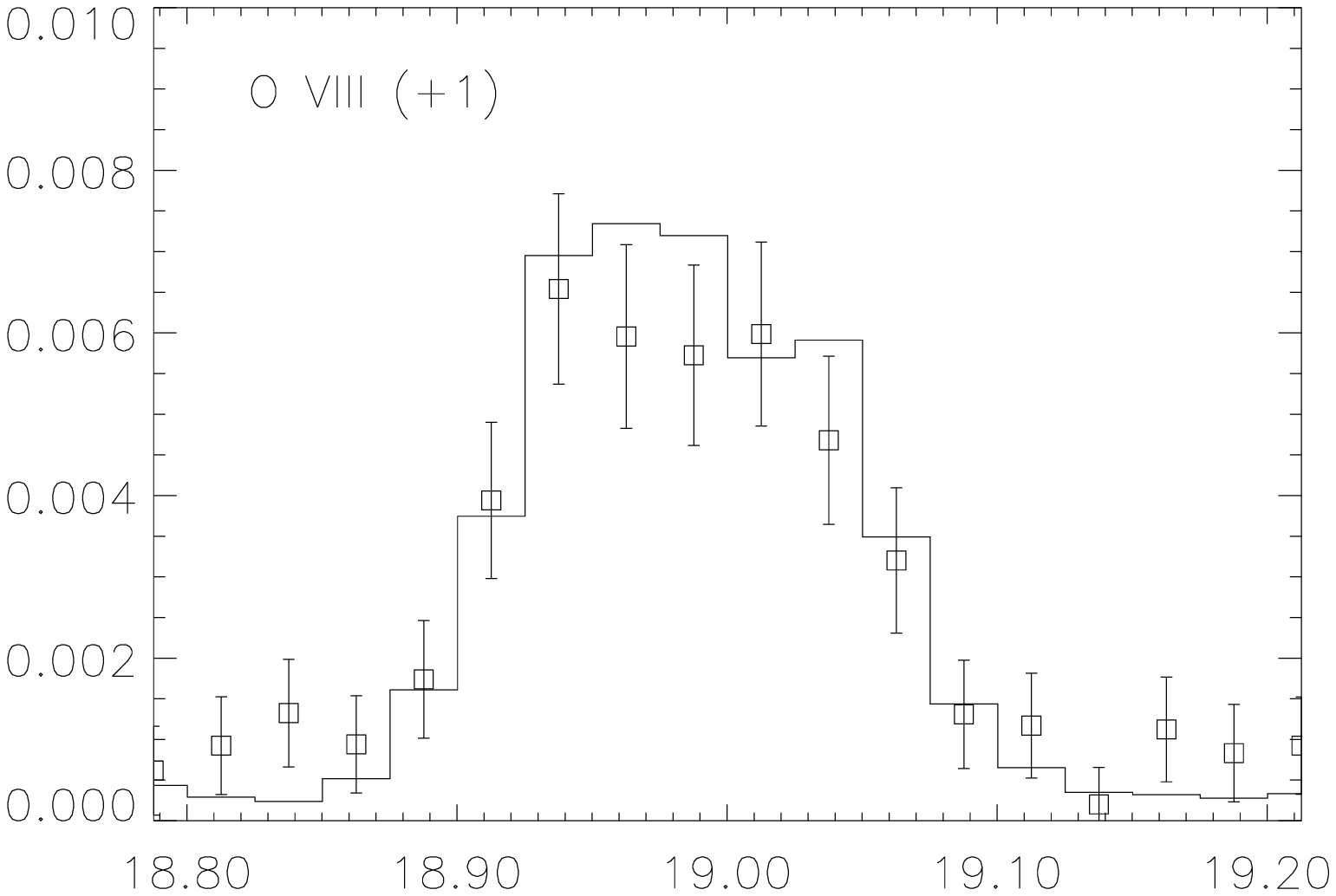} 
\includegraphics[width=3in, height=1.85in]{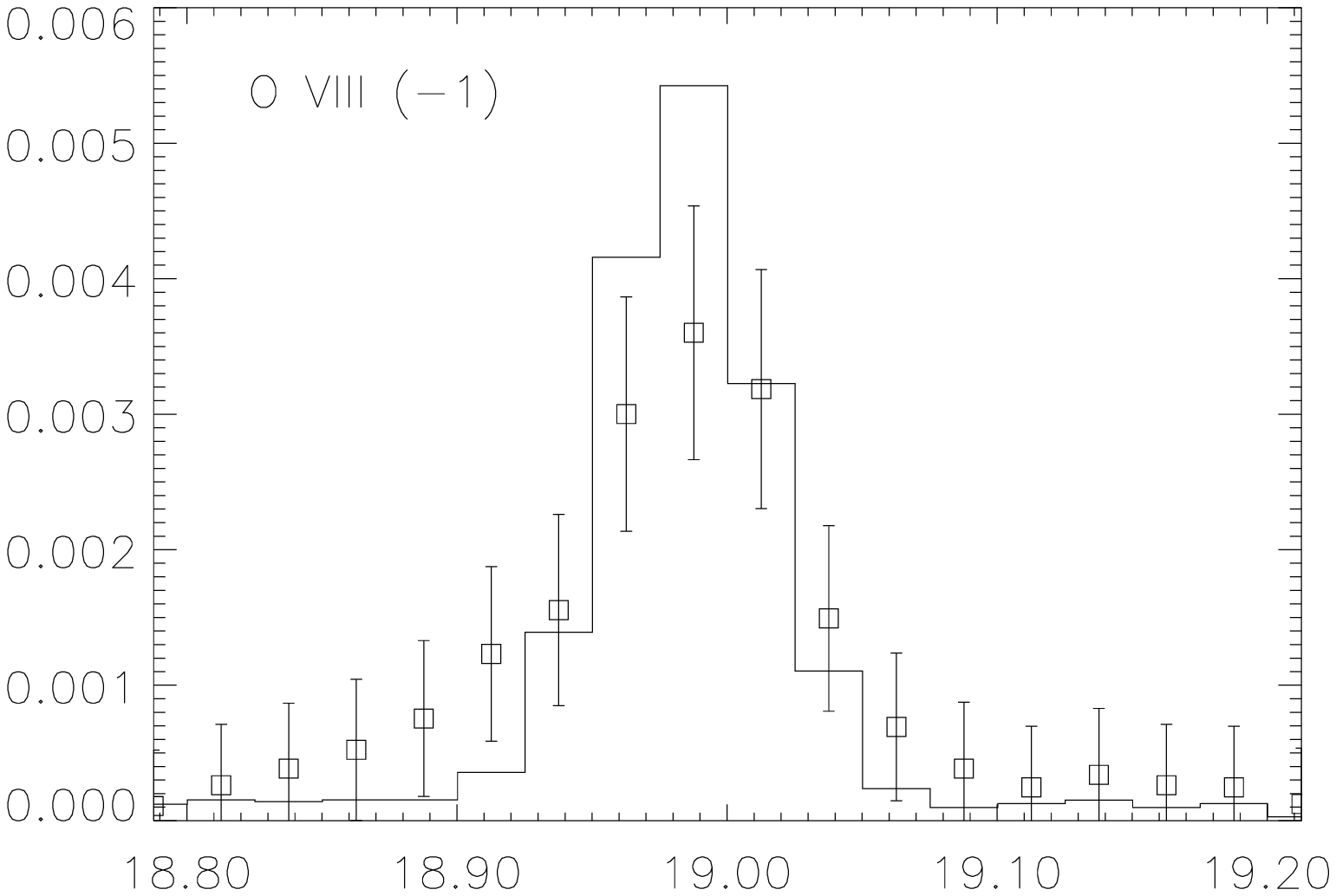} 
\includegraphics[width=3in, height=1.85in]{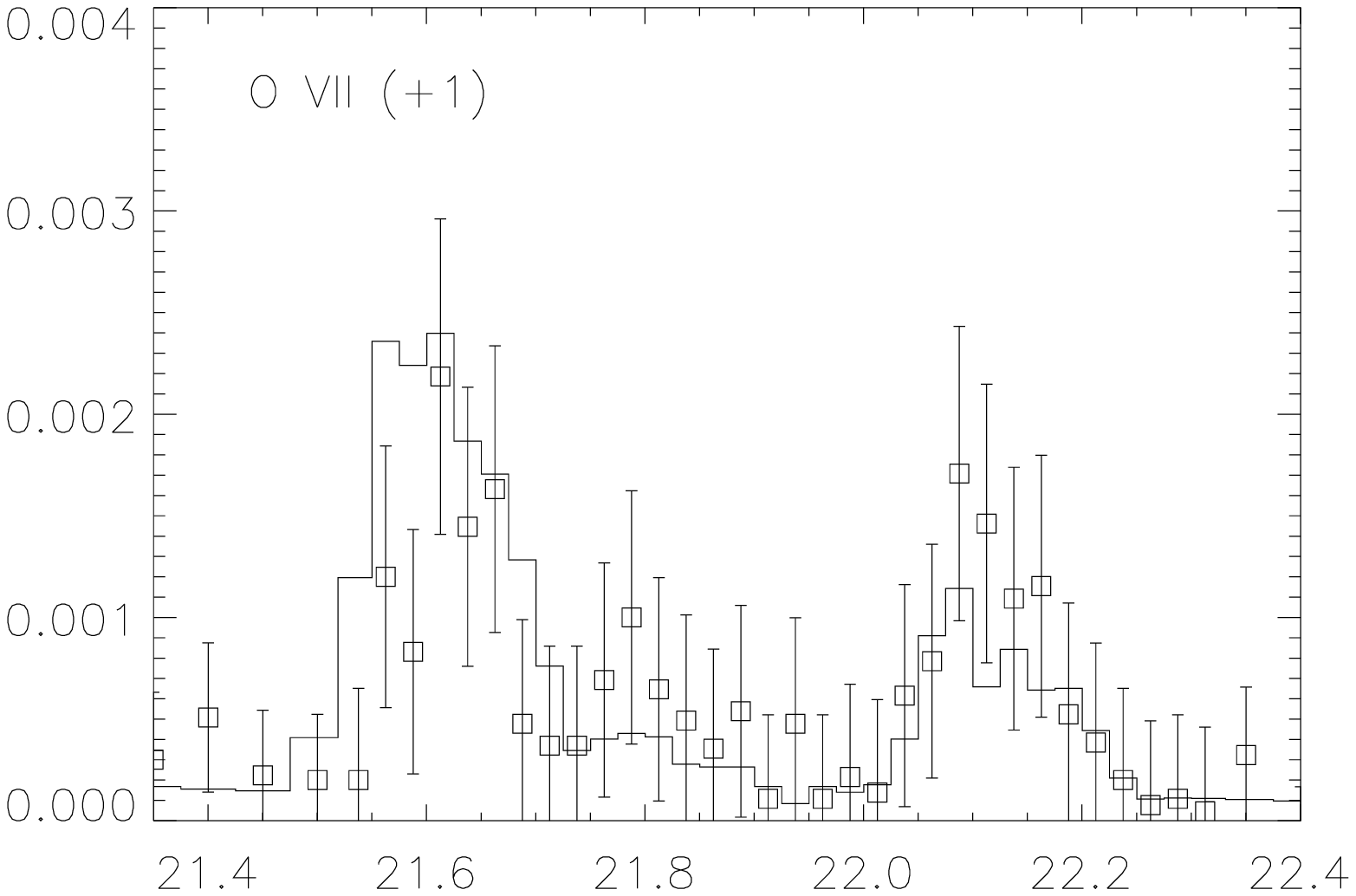} 
\includegraphics[width=3in, height=1.85in]{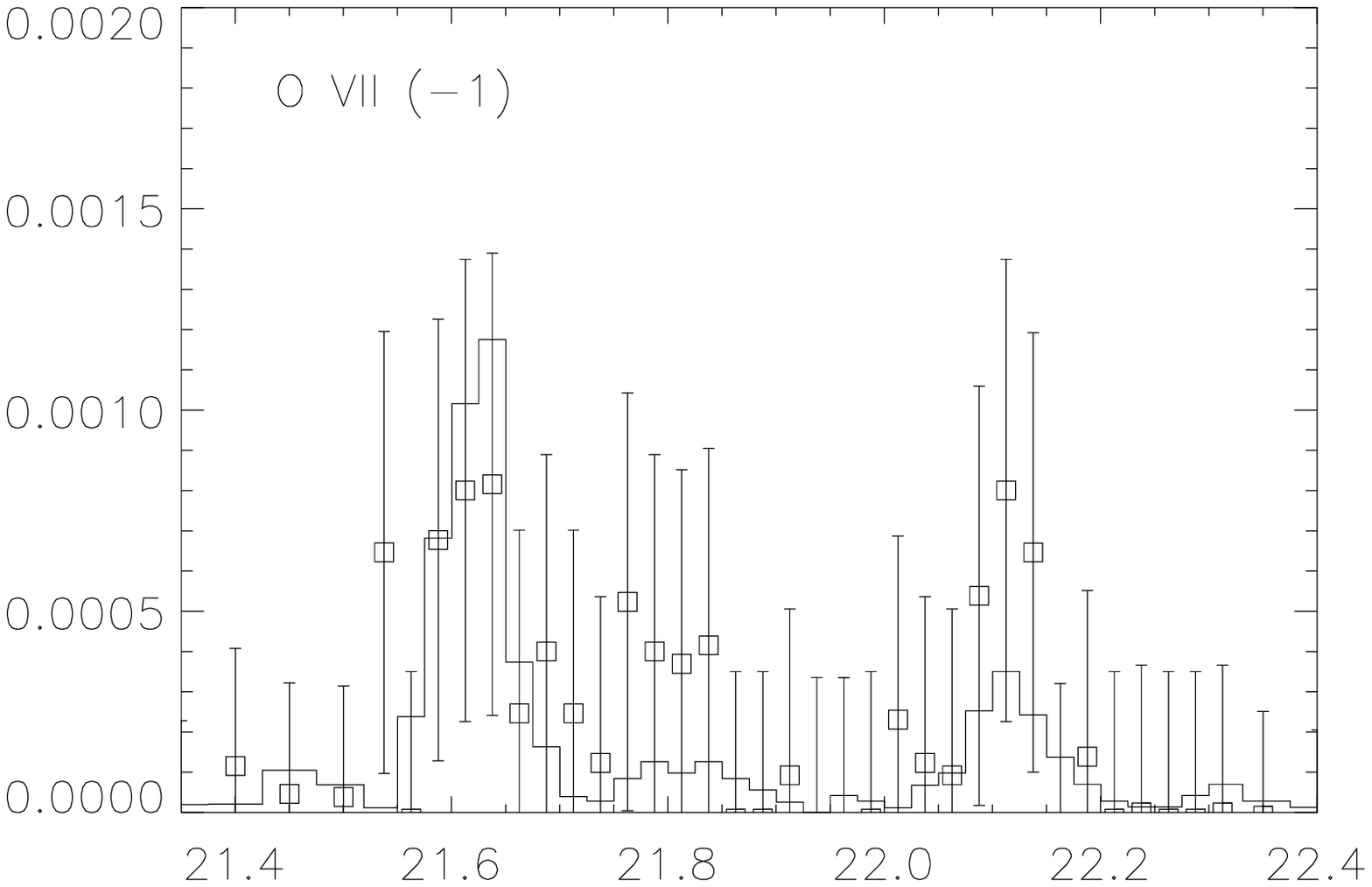} 
\includegraphics[width=3in, height=1.85in]{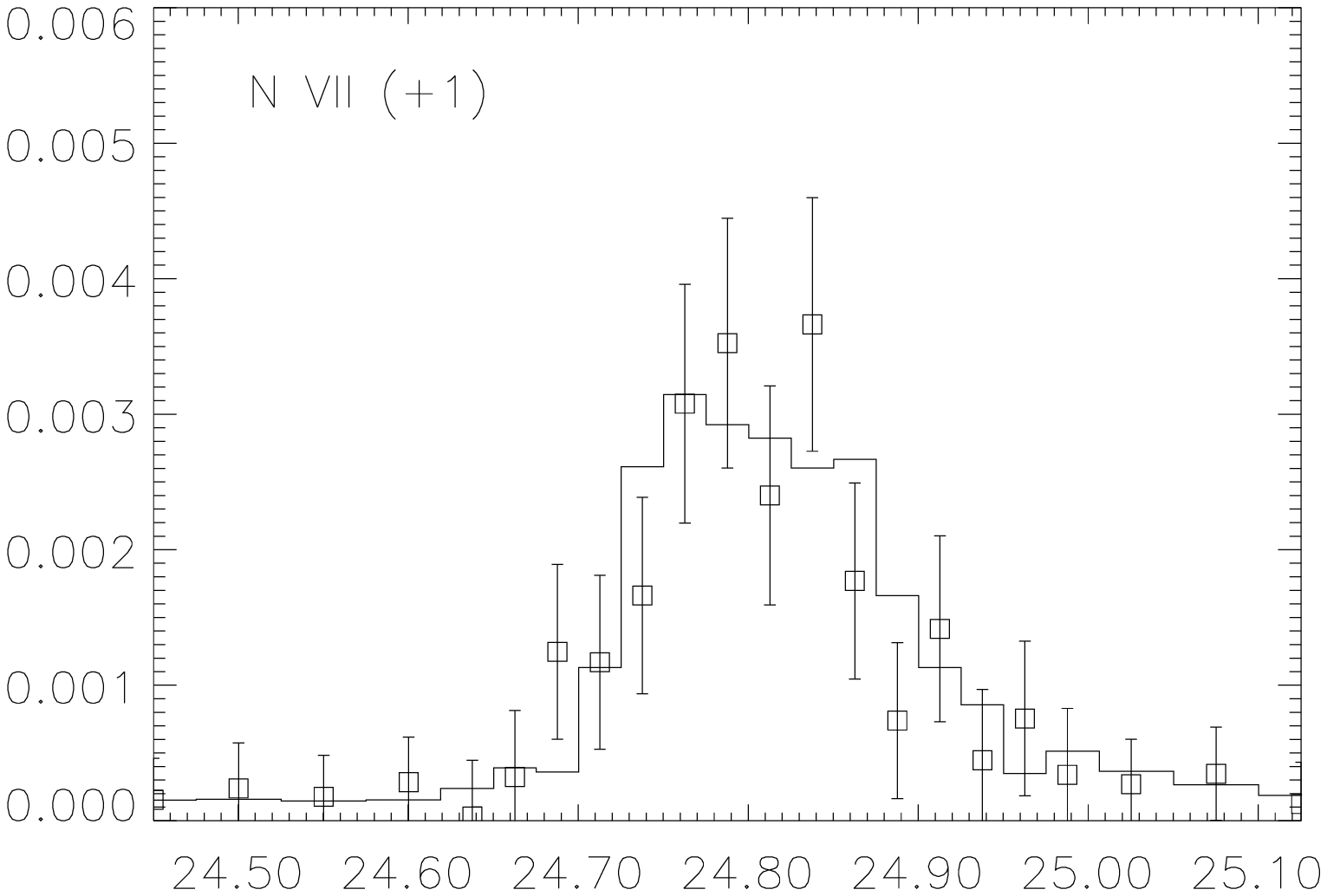} 
\includegraphics[width=3in, height=1.85in]{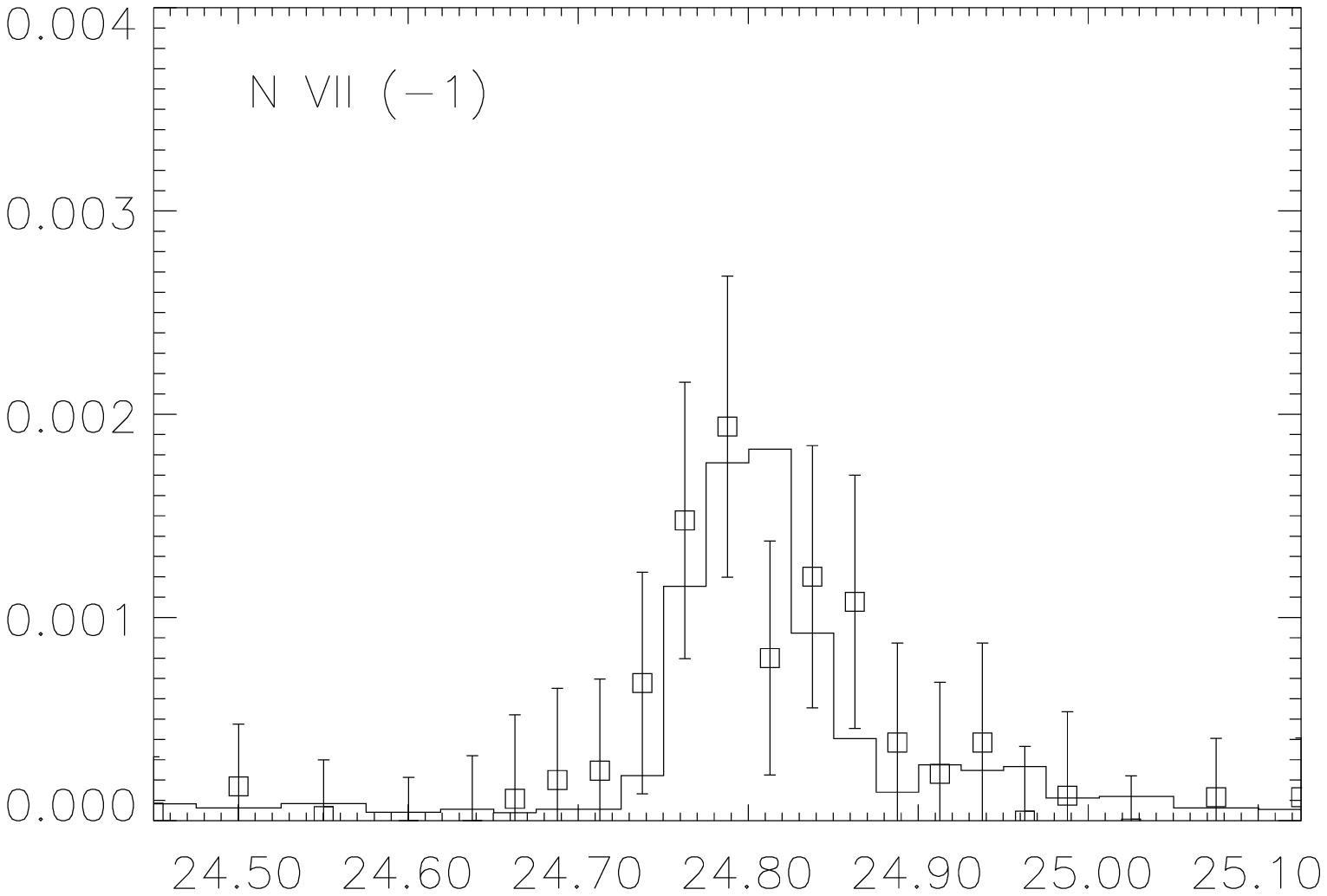} 
\end{center} 
\caption{ The same as in
Fig.~\ref{fig:marx1} but for Fe XVII, the H-like and He-like ions iof
O VIII (Ly$_{\alpha}$) and O VII and the H-like ion of N VII.  
}
\label{fig:marx3} 
\end{figure}

Figures \ref{fig:marx1}, \ref{fig:marx2} and \ref{fig:marx3} display
these average spectra overlaid with the observed positive and negative
first order LETG spectra.  The excellent correspondence between the
simulated and actual spectra gives us confidence that our model is a
reasonable approximation to the actual interaction of the blast wave
with the inner ring.

\section{Discussion and Conclusions} 
\label{sec:disc} 

The analysis of the LETG spectra of SNR~1987A presented in \S
\ref{sec:global} and \S \ref{sec:marx} of this paper confirms and
refines the main conclusions of Paper I.  The X-ray emitting plasma is
largely confined to a radially expanding ring near the inner optical
ring.  The X-ray emitting gas is heated by a distribution of shocks
and is not in ionization equilibrium. The electron temperature
distribution of the shocked gas is bimodal, with peaks at $\sim 0.5$
and $\sim 3$~keV, respectively.

Our analysis leads us to a general picture of SNR~1987A in which the
X-ray emission is dominated by two components.  The hot ($kT_e \sim
3$~keV) component results from shocks of relatively high velocity that
propagate through relatively low density $(n_e \sim 10^2$ cm$^{-3}$)
circumstellar gas.  The cool ($kT_e \sim 0.5$~keV) component results
from shocks transmitted into the denser $(n_e \sim 10^4$ cm$^{-3}$)
gas of the inner circumstellar ring.  

\begin{figure} 
\begin{center} 
\includegraphics[width=5in, height=3in]{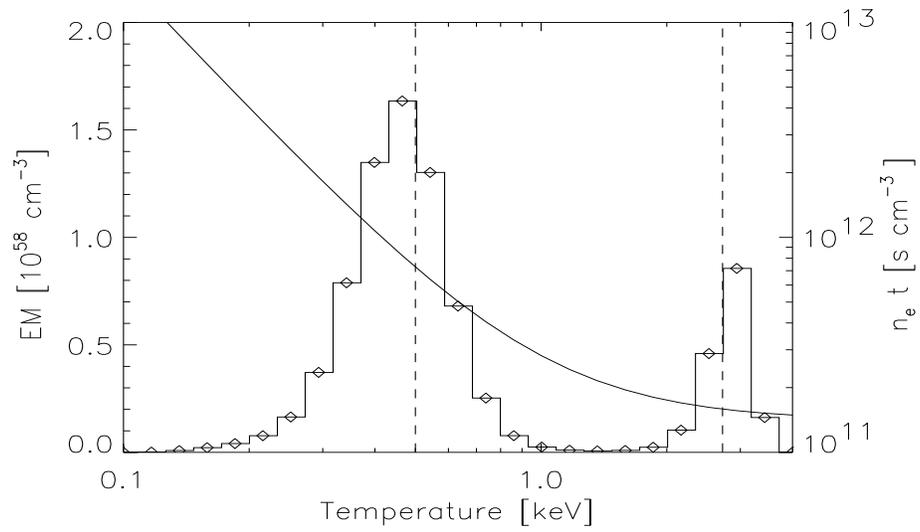} 
\end{center} 
\caption{ Emission measure (EM)
distribution of shocks in SNR~1987A as derived from the DS model with
25 points logarithmically spaced in the (0.1 - 4 keV) postshock
temperature range. The two vertical dashed lines indicate the plasma
temperature derived from the discrete two-shock model. The solid line
shows the derived ionization age of each shock ($n_e t$).
} 
\label{fig:DS} 
\end{figure}

The fact that the inferred ionization age decreases with shock
velocity (Fig.~\ref{fig:DS}) is consistent with this picture.  The
slower shocks propagate through the higher density gas, while the
faster shocks propagate through the lower density gas.  Therefore, if
the typical timescale since the shock has entered the gas is
comparable ($t \sim$~ few years), the ionization age, $n_et$, of the
gas behind the slower shocks will be substantially greater than that
of the gas behind the faster shocks.  

In their analysis of a spectrum of SNR~1987A taken with the High
Energy Transmission Grating of Chandra in October 1999, Michael et al.
(2002) inferred an electron temperature, kT$_e \sim 2.6$~keV from the
line ratios.  By stacking all the emission lines into a composite
profile, they also a measured the width, FWHM $\approx 5000$\kms, of a
composite line profile, from which they inferred a `typical' shock
velocity $\approx 3500$\kms that was consistent with the radial
expansion velocity of the X-ray image as measured by Park et al.
(2002). 

Since the gas behind such a shock would have electron temperature
$\sim 17$ keV if the electrons and ions were in temperature
equilibrium, Michael et al. concluded that the electron and ion
temperatures had not equilibrated.  This was a reasonable scenario,
since the electron-ion temperature equilibration time, $(n_et) =
4\times10^{11}(kT_{keV})^{1.5}$~cm$^{-3}$~s (Spitzer 1962), of such a
gas would be substantially greater than the typical ionization age,
$n_et \sim 4 \times 10^{10}$~cm$^{-3}$~s, of shocked gas having $n_e
\approx 400$~cm$^{-3}$ and $t \sim 3$ years. 

But the shock environment of SNR~1987A has changed considerably since
that observation.  The line profiles of the current (September 2004)
observations (Paper I) indicate shock velocities $300-1700$\kms, and
hence typical post-shock temperatures kT$_e ~ \sim 0.1 - 4$~keV.
Moreover, such low shock velocities suggest that the shocked gas must
have greater density, say, $4 \times 10^4 > n_e > 1600$~cm$^{-3}$.
From these considerations and Fig.~\ref{fig:DS}, we conclude that
there is no longer any compelling evidence that the electrons and ions
in the shocked gas are out of temperature equilibrium, although this
may still be the case for gas behind the faster shocks propagating
through the lower density gas.

We expect the rapid evolution of the X-ray spectrum to continue. Park
et al. (2005) found that the X-ray light curve is brightening at an
accelerating rate.  This fact and the correlation of the X-ray image
with the optical hotspots (Park et al. 2006, in preparation)
indicate that the blast wave is interacting
with much denser gas than in the past. This observation is supported
by the noticeable deceleration of the radial expansion velocity of the
X-ray image.  The most recent {\it Chandra} data show that this
velocity has decreased from $\sim 4000$\kms (Park et al. 2004)
to $\sim 1600$\kms (Racusin et al. 2006, in
preparation).  

The rapid brightening of the X-ray light curve of SNR~1987A (Park et
al. 2005) ensures that we may soon be able to obtain spectra with
excellent photon statistics.  Thus, with future {\it Chandra}
observations we should be able to constrain the model further and
subject it to some critical tests.  

When the blast wave strikes dense clumps of gas in the circumstellar
ring, the interaction will give rise to both transmitted and reflected
shocks, both of which will cause enhanced X-ray emission correlated
with the optical hotspots.  In both cases the velocity distribution of
the shocked gas will shift toward lower velocities.  Therefore, we
expect that the X-ray emission lines will continue to become narrower.  

The evolution of the temperature distribution of the X-ray emitting
gas will tell us the relative contribution of reflected and
transmitted shocks to the rapidly brightening source.  If the gas
behind the transmitted shocks dominates the X-ray emission, the
bimodal temperature distribution of the X-ray emitting gas will become
more skewed towards lower temperatures. In fact, Park et al. (2005)
have already noticed this trend in their two-shock model fits to {\it
Chandra} nondispersed (ACIS-S) spectra over the past five years.  On
the other hand, if the gas behind the reflected shocks contributes
substantially to the X-ray emission, the hot component in the X-ray
spectrum will continue to remain substantial.  

The radio emission from SNR~1987A is non-thermal (Gaensler et al.
1997; Manchester et al. 2002) and is probably produced by relativistic
electrons accelerated by the reverse shock (Manchester et al. 2005).
Park et al. (2005) have noted that the hard X-ray light curve
resembles the radio light curve of SNR~1987A and have suggested that
the radio emission and the hard X-rays might have a common origin.
This suggestion raises the question of whether the X-ray spectrum may
have a non-thermal (i.e., power-law) component.  Up to now, we have
seen no compelling evidence for the presence of such a component,
either in the pulse-height spectra (Michael et al. 2002) or in the LETG
spectrum discussed here.  On the other hand, we cannot rule out the
possibility of a significant non-thermal component, especially at
photon energies $> 4$~keV.  

Observations of broad line profiles of L$_{\alpha}$ and H$_{\alpha}$
(Michael et al. 2003; Smith et al. 2005) show that the supernova
debris is crossing this reverse shock with velocities $\sim 10^4$\kms.
If the hard X-rays originate from the region behind the reverse shock
and are produced by thermal processes, then we would expect to see
strong emission lines of high-ionization species with relatively broad
profiles ($\sim$ few $10^3$\kms).  On the other hand, if the hard
X-rays are predominantly non-thermal, we would expect to see a
decrease in the equivalent widths of the X-ray emission lines at
higher energy.  

Future {\it Chandra} observartions will be also helpful in
resolving the issue with the CNO abundances derived in X-rays. 
As discussed in \S~\ref{sec:global} (see also Table~\ref{tab:fit}),
the N/O ratio is consistent with the value derived from the 
analysis of the optical/UV spectra of the inner ring in SNR~1987A,
but the total amount of light metals (C, N, O)
derived in our analysis (CNO $\approx 0.15$~solar) is a factor of two lower
than that derived from optical/UV spectra (CNO $\approx 0.28$~solar, 
see Lundqvist \& Fransson 1996). The latter ratio is consistent with
the average CNO abundances for LMC (CNO $\approx 0.26$~solar,
Russel \& Dopita 1992).  This means that the X-ray analysis 
confirms the enhancement of the N/O ratio, which may be attributed to 
CNO processing of the matter by the progenitor star.  But the X-ray 
observations 
indicate a deficit in the net abundance of C $+$ N $+$ O (the CNO processing 
cannot alter their total amount). 
On the other hand, the abundances of 
heavier metals (Ne, Mg, Si, S and Fe) derived here are consistent
with typical values for the SNRs in LMC (Hughes et al. 1998).

What could account for the low abundances of C $+$ N $+$ O derived 
from the X-ray spectrum?  On the one hand, we cannot rule out the 
possibility of purely technical reasons for this discrepancy, such 
as greater uncertainties in the X-ray spectrum at low energies due 
to the CCD degradation or uncertainties in the atomic data.  
On the 
other hand, there might be physical reasons.  Could the amount of CNO 
elements in the progenitor star be considerably lower (a factor of two) 
than the average for LMC?  Or, could there be an extra source of 
(likely nonthermal) continuum, the effect of which would be to reduce 
the CNO abundances if interpreted as thermal continuum?  Future 
{\it Chandra} grating observations with anticipated better photon 
statistics will help us find the correct answer. In addition, 
such observations may also reveal some evolution (increase) for the 
abundances of
elements like O, Mg, Si and Fe resulting from the destruction of dust
from the supernovae debris and the inner circumstellar ring
by the shocked  hot gas emitting in X-rays.





\acknowledgments This work was supported by NASA through Chandra
Awards G04-5072A (to CU, Boulder, CO) and GO4-5072B (to NCSU, Raleigh,
NC).

\clearpage






%
\begin{table}
\begin{center}
\caption{Shock Fitting Results (1-st and 0-th order spectra)
\label{tab:fit}}
\begin{tabular}{llll}
\tableline
\tableline
\multicolumn{1}{l}{Parameter} &\multicolumn{2}{c}{LETG 1-st}  &
\multicolumn{1}{c}{LETG 0-th}  \\
\tableline
          & 1-shock & 2-shock  & 2-shock  \\
\tableline
$\chi^2$/dof  & 661/436 & 417/433 &  121/111   \\
N$_H$(10$^{21}$ cm$^{-2}$) & 0.63 [0.45 - 0.82] & 1.47 [1.25 - 1.73] &
2.13 [1.52 - 3.00] \\
kT$_1$ (keV)  &  & 0.50 [0.47 - 0.54]& 0.41 [ 0.35 - 0.66] \\
kT$_2$ (keV)  & 2.16 [2.08 - 2.21] & 2.72 [2.44 - 3.02] & 3.40 [2.63 -
4.06] \\
tau$_1$$^{(a)}$  &  & 5.21 [3.73 - 7.83] & 6.36 [3.10 - 29.1] \\
tau$_2$$^{(a)}$  & 0.69 [0.62 - 0.77] & 1.43 [0.77 - 15.4] &
1.43 [0.76 - 13.5]\\
EM$_1$$^{(b)}$      &  & 5.51 &  12.7 \\
EM$_2$$^{(b)}$      & 2.45 & 1.69 & 1.44 \\
H    & 1 & 1 & 1 \\
He(2.57)        & 2.57 & 2.57 & 2.57  \\
C~(0.09)   & 0.09 & 0.09 & 0.09  \\
N~(1.63)   & 0.40 [0.32 - 0.50] & 0.77 [0.60 - 1.00] & 0.12 [0.0 -
0.73]\\
O~(0.18)   & 0.055 [0.047 - 0.066] & 0.092 [0.075 - 0.104] & 0.10
[0.073 -
0.15]\\
Ne(0.29)   & 0.25 [0.22 - 0.28] & 0.29 [0.25 - 0.34] & 0.29 [0.21 -
0.41] \\
Mg(0.32)   & 0.21 [0.18 - 0.24] & 0.24 [0.20 - 0.28] & 0.23 [0.16 -
0.33] \\
Si(0.31)   & 0.24 [0.21 - 0.27] & 0.28 [0.22 - 0.32] & 0.66 [0.47 -
1.12] \\
S~(0.36)   & 0.42 [0.28 - 0.56] & 0.45 [0.32 - 0.60] & 0.43 [0.24 -
0.96] \\
Ar(0.54) & 0.54 & 0.54 & 0.54 \\
Ca(0.34) & 0.34 & 0.34 & 0.34 \\
Fe(0.22)   & 0.16 [0.14 - 0.18] & 0.16 [0.14 - 0.17] & 0.11 [0.08 -
0.18] \\
Ni(0.62) & 0.62 & 0.62 & 0.62 \\
F$^{(c)}_X$(0.5-2 keV) &
          1.42 & 1.47  &  1.55  \\
F$^{(c)}_X$(0.5-6 keV) &
          1.80 & 1.84  &  1.95  \\
\tableline
\tableline
\end{tabular}

\tablenotetext{}{ The 90\%-confidence intervals are given in brackets
and all abundances are expressed as ratios to their solar values
(Anders \& Grevesse 1989).  For comparison, the inner-ring abundances
of He, C, N, and O (Lundqvist \& Fransson 1996); those of Ne, Mg, Si, S,
and Fe typical for the LMC SNRs (Hughes et al. 1998); and 
Ar, Ca and Ni abundances  representative for LMC (Russell \& Dopita 1992)
are given in the first column in parentheses. Note that the H, He, C,
Ar, Ca, and Ni abundances were kept fixed in all model fits (see text).
} 
\tablenotetext{a}{ ($n_e t$) in units of $10^{11}$ s cm$^{-3}$.} 
\tablenotetext{b}{ $EM = \int n_e n_H dV$ in units of
$10^{58}$~cm$^{-3}$, assuming a distance of 50 kpc.}
\tablenotetext{c}{X-ray flux in units of $10^{-12}$ ergs cm$^{-2}$
s$^{-1}$.} 
\end{center} 
\end{table} %

\clearpage

\clearpage 

\clearpage

\clearpage

\clearpage

\clearpage

\clearpage

\end{document}